\documentclass[journal]{IEEEtran}
\usepackage[T1]{fontenc}
\usepackage[tableposition=top]{caption}
\usepackage{url}

\usepackage{breakurl}
\usepackage[breaklinks]{hyperref}
\usepackage{rotating}
\usepackage{color, colortbl}
\usepackage{booktabs, array, dcolumn}
\usepackage{tikz}
\usepackage{graphicx}
\usepackage{pgf-umlsd}
\usepackage{multirow}
\usepackage{epsfig}
\usepackage{subfig}
\usepackage{pgfplotstable}
\usepackage{amsfonts}
\usepackage{relsize}
\usepackage{varwidth}
\usepackage{amsmath}
\usepackage{amssymb}
\usepackage{booktabs}
\usepackage[normalem]{ulem}
\usepackage{subfig}
\usepackage{kernelstyle}
\definecolor{Gray}{gray}{0.9}
\usepackage{footnote}
\usepackage{longtable}
\usepackage{multirow}
\usepackage{siunitx}
\usepackage{array,graphicx}
\usepackage{booktabs}
\usepackage{pifont}
\usepackage{bm}

\newcommand*\rot{\rotatebox{45}}
\newcommand*\rott{\rotatebox{90}}

\definecolor{gray1}{gray}{0.9}
\definecolor{gray2}{gray}{0.6}
\definecolor{Gray}{gray}{0.9}

\definecolor{light-gray1}{gray}{0.65}
\definecolor{light-gray2}{gray}{0.95}

\DeclareCaptionFormat{myformat}{%
  \begin
  {varwidth}{\linewidth}%
    \centering
    #1#2#3%
  \end{varwidth}%
}

\newcommand\copyrighttext{%
  \footnotesize This work has been submitted to the IEEE for possible publication. Copyright may be transferred without notice, after which this version may no longer be accessible}
\newcommand\copyrightnotice{%
\begin{tikzpicture}[remember picture,overlay]
\node[anchor=south,yshift=10pt] at (current page.south) {\fbox{\parbox{\dimexpr\textwidth-\fboxsep-\fboxrule\relax}{\copyrighttext}}};
\end{tikzpicture}%
}

\begin{document}

\title{A Multi-modal Neural Embeddings  Approach for Detecting Mobile Counterfeit Apps: A Case Study on Google Play Store}

\author{Naveen Karunanayake,
        Jathushan Rajasegaran, 
	Ashanie Gunathillake,\\
	Suranga Seneviratne, and 
	Guillaume Jourjon 
\thanks{This is an extension of our previous work~\cite{rajasegaran2019multi} published in the The Web Conference (WWW), 2019.}%
\thanks{N. Karunanayake is with the CSIRO-Data61, Australia and University of Moratuwa Sri Lanka.}
\thanks{J. Rajasegaran, and G. Jourjon are with the CSIRO-Data61, Australia.}
\thanks{A. Gunathillake and S. Seneviratne are with the School of Computer Science, The University Sydney, Australia.}}

\markboth{Transactions on Mobile Computing,~Vol.~XX, No.~XX, 20XX}%
{Rajasegaran \MakeLowercase{\textit{et al.}}: A Multi-modal Neural Embeddings  Approach for Detecting Mobile Counterfeit Apps}


\IEEEtitleabstractindextext{
\begin{abstract}
Counterfeit apps impersonate existing popular apps in attempts to misguide users to install them for various reasons such as collecting personal information, spreading malware, or simply to increase their advertisement revenue. Many counterfeits can be identified once installed, however even a tech-savvy user may struggle to detect them before installation as app icons and descriptions can be quite similar to the original app. To this end, this paper proposes to leverage the recent advances in deep learning methods to create image and text embeddings so that counterfeit apps can be efficiently identified when they are submitted to be published in app markets. We  show that for the problem of counterfeit detection, a novel approach of combining \emph{content embeddings} and \emph{style embeddings} (given by the Gram matrix of CNN feature maps) outperforms the baseline methods for image similarity such as SIFT, SURF, LATCH, and various image hashing methods. We first evaluate the performance of the proposed method on two well-known datasets for evaluating image similarity methods and show that, content, style, and combined embeddings increase \emph{precision@k} and \emph{recall@k} by 10\%-15\% and 12\%-25\%, respectively when retrieving five nearest neighbours. Second specifically for the app counterfeit detection problem, combined content and style embeddings achieve 12\% and 14\% increase in \emph{precision@k} and \emph{recall@k}, respectively compared to the baseline methods. We also show that adding text embeddings further increases the performance by 5\% and 6\% in terms of \emph{precision@k} and \emph{recall@k}, respectively when $k$ is five. Third, we present an analysis of approximately 1.2 million apps from Google Play Store and identify a set of potential counterfeits for top-10,000 popular apps. Under a conservative assumption, we were able to find 2,040 potential counterfeits that contain malware in a set of 49,608 apps that showed high  similarity to one of the top-10,000 popular apps in Google Play Store. We also find 1,565 potential counterfeits asking for at least five additional dangerous permissions than the original app and 1,407 potential counterfeits having at least five extra third party advertisement libraries.



\end{abstract}

%
%
%

}

\maketitle

\IEEEdisplaynontitleabstractindextext

\IEEEpeerreviewmaketitle

\copyrightnotice
\section{Introduction}
\label{Sec:Introduction}

Availability of third party apps is one of the major reasons behind the wide adoption of smartphones. The two most popular app markets, Google Play Store and Apple App Store, hosted \textcolor{black}{approximately 2.85 million and 1.8 million apps at the first quarter of 2020~\cite{google2020,apple2020} and these numbers are likely to grow further.} Handling such large numbers of apps is challenging for app market operators since there is always a trade-off between how much scrutiny is put into checking apps and encouraging developers by providing fast time-to-market. As a result, problematic apps of various kinds have made it into the app markets, including malware, before being taken down after receiving users' complaints~\cite{zhou2012dissecting,Seneviratne2015}. 

\begin{figure}[h]
\centering
\includegraphics[scale=0.3]{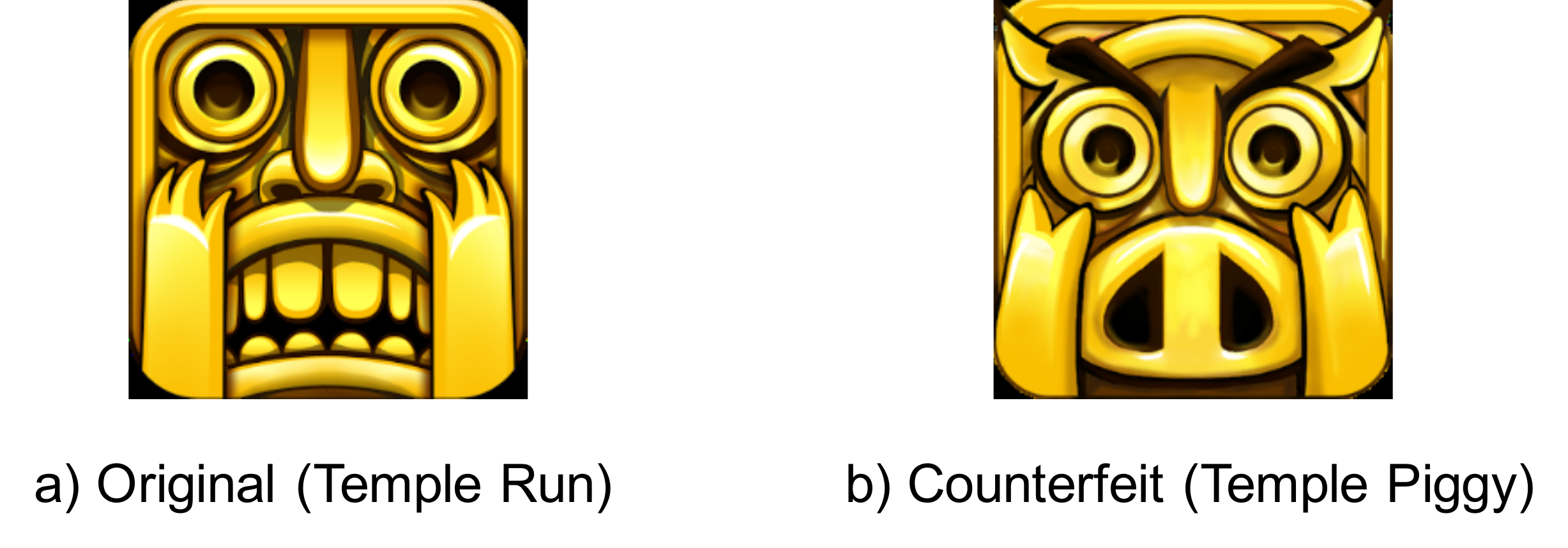}%
\caption{An example counterfeit app for the popular arcade game Temple Run} 
\label{Fig:Ex1}
\end{figure}

One category of problematic apps making into app markets is \emph{counterfeits} (i.e. apps that attempt to impersonate popular apps). The overarching goals behind app impersonation can be broadly categorised into two. First, developers of counterfeits are trying to attract app installations and increase their advertisement revenue. This is exacerbated by the fact that some popular apps are not available in some countries and users who search the names of those popular apps can become easy targets of impersonations. Second is to use counterfeits as a means of spreading malware. For instance, in November 2017 a fake version of the popular messenger app \emph{WhatsApp}~\cite{whatsapp2017} was able to get into Google Play Store and was downloaded over 1 million times before it was taken down. Similar instances were reported in the past for popular apps such as \emph{Netflix}, \emph{IFTTT}, \emph{Angry Birds}~\cite{angrybirds2012,sarah2013new,netflix2017}, and \emph{banking apps}~\cite{cba2018}. More recently, counterfeits have been used to secretly mine crypto currencies in smartphones~\cite{securelist2018}. Unlike the \emph{app clones}~\cite{crussell2013andarwin} that show code level similarity, these counterfeits show no similarity in codes and only appearance-wise similar to the original app. In \figurename~\ref{Fig:Ex1}, we show an example counterfeit named \emph{Temple Piggy}\footnote{\emph{Temple Piggy} is currently not available in Google Play Store.}  which shows a high visual similarity to the popular arcade game \emph{Temple Run}~\cite{TempleRun}.



%
%
%


In this paper, we propose a neural embedding-based apps similarity detection framework that allows to identify counterfeit apps from a large corpus of apps represented by their icons and text descriptions. We leverage the recent advances in Convolutional Neural Networks (CNNs) to generate feature embeddings from given images using pre-trained models such as AlexNet~\cite{alexNet}, VGGNet~\cite{vggNet}, and ResNet~\cite{resNet}. However, in contrast to commonly used \emph{content embeddings} generated from fully connected layers before the last soft-max layer, we show that combining \emph{content embeddings} with \emph{style embeddings} generated from the Gram matrix of convolutional layer feature maps of the same pre-trained models achieve better results in detecting visually similar app icons. Specifically, following are the main contributions of this paper.




\begin{itemize} 

\item We show that the novel method of using combined \emph{style} and \emph{content} embeddings generated from pre-trained CNNs outperforms many baseline image retrieval methods including hashing, feature-based, and structural similarity based methods such as SIFT~\cite{SIFT}, SURF~\cite{SURF}, and SSIM~\cite{SSIM}, for the task of detecting visually similar app icons. We also validate this method using two standard image retrieval datasets; \emph{Holidays dataset}~\cite{jegou2008hamming} and \emph{UKBench}~\cite{UKBench}, and show that neural embeddings also perform better than baseline hashing and feature-based methods.



\item Using a large dataset of over 1.2 million app icons, we show that \emph{combined content and style embeddings} achieve 8\%--12\% higher \emph{precision@k} and 14\%--18\% higher \emph{recall@k} when $k\in\{5,10\}$.

\item We show that adding \emph{text embeddings}~\cite{le2014distributed} generated using the app description as an additional modality for similarity search, further increases the performance by 3\%--5\% and 6\%--7\% in terms of \emph{precision@k} and \emph{recall@k}, respectively when $k\in\{5,10\}$.



\item We identify a set of  7,246 potential counterfeits (that are similar both visually as well as functionality-wise) to the top-10,000 popular apps in Google Play and show that under a conservative assumption, 2,040 of them contain malware.  We further show that 1,565 potential counterfeits ask for at least five additional dangerous permissions than the original app and 1,407 potential counterfeits have at least five extra third party advertisement libraries. To the best of our knowledge this is the first large-scale study that investigates the depth of app counterfeit problem in app stores.
\end{itemize} 


The remainder of the paper is organised as follows. In Section~\ref{Sec:RelatedWork}, we present the related work. In Section~\ref{Sec:Dataset}, we introduce our dataset followed by the methodology in Section~\ref{Sec:Methodology}. \textcolor{black}{Section~\ref{Sec:Results} presents our results, while Section~\ref{Sec:Discussion} discusses implications of our findings, limitations, and possible future extensions.}  Section~\ref{Sec:Conclusion} concludes the paper. 

\section{Related Work}
\label{Sec:RelatedWork}

\subsection{Mobile Malware \& Greyware}
While there is a plethora of work on detecting mobile malware~\cite{grace2012riskranker,wu2012droidmat,burguera2011crowdroid,shabtai2012andromaly,yuan2014droid} and various fraudulent activities in app markets~\cite{xie2015appwatcher,chandy2012identifying,gibler2013adrob,surian2017app,seneviratne2017spam,Seneviratne2015}, only a limited amount of work focused on the \emph{similarity of mobile apps}. One line of such work focused on detecting clones and rebranding. Viennot et al. ~\cite{viennot2014measurement} used the Jaccard similarity of app resources in the likes of images and layout XMLs to identify clusters of similar apps and then used the developer name and certificate information to differentiate clones from rebranding. Crussell et al.~\cite{crussell2013andarwin} proposed to use features generated from the source codes to identify similar apps and then used the developer information to isolate true clones. In contrast to above work, our work focuses on identifying visually similar apps (i.e. counterfeits) rather than the exact similarity (i.e. clones), which is a more challenging problem.

Limited amount of work focused on identifying visually similar mobile apps~\cite{sun2015droideagle,malisa2016mobile,andow2016study,Malisa:2017}. For example, Sun et al.~\cite{sun2015droideagle} proposed DroidEagle that identifies the visually similar apps based on the LayoutTree of XML layouts of Android apps. While the results are interesting this method has several limitations. First, all visually similar apps may not be necessarily similar in XML layouts and it is necessary to consider the similarities in images. Second, app developers are starting to use code encryption methods, thus accessing codes and layout files may not always possible. Third, dependency of specific aspects related to one operating system will not allow to make comparisons between heterogeneous app markets and in such situations only metadata and image similarity are meaningful. Recently, Malisa et al.~\cite{Malisa:2017} studied how likely would users detect a spoofing application using a complete rendering of the application itself. To do so, authors introduced a new metric representing the distance between the original app screenshot and the spoofing app. In contrast to above work, the proposed work intends to use different neural embeddings derived from app icons  and text descriptions that will better capture visual and functional similarities.



\subsection{Visual Similarity \& Style Search}

Number of work looked into the possibility of transferring style of an image to another using neural networks. For example, Gatys et al.~\cite{gatys2015neural,gatys2016image} proposed a \emph{neural style transfer algorithm} that is able to transfer the stylistic features of well-known artworks to target images using feature representations learned by Convolutional Neural Networks (CNNs). Several other methods proposed to achieve the same objective either by updating pixels in the image iteratively or by optimising a generative model iteratively and producing the styled image through a single forward pass. A summary of the available style transfer algorithms can be found in the survey by Jing et al.~\cite{jing2017neural}.

Johnson et al.~\cite{johnson2016perceptual} have proposed a feed-forward network architecture capable of real-time style transfer by solving the optimisation problem formulated by Gatys et al.~\cite{gatys2016image}. Similarly, to style transfer, CNNs have been successfully used for image searching. In particular, Bell~\&~Bala~\cite{bell2015siamese} proposed a Siamese CNN to learn a high-quality embedding that represent visual similarity and demonstrated the utility of these embeddings on several visual search tasks such as searching products across or within categories. Tan et al.~\cite{tan2016ceci} and  Matsuo \& Yanai~\cite{style_classification} used embeddings created from CNNs to classify artistic styles. 

This paper is an extension of our previous work~\cite{rajasegaran2019multi} where we demonstrated the feasibility of the multi-modal embeddings for app counterfeit detection. In this paper we expand our analysis further \textcolor{black}{by assessing the impact of different embeddings and pre-trained models and provide a detailed view on multiple motivations behind app counterfeiting. To the best of our understanding, our work is the first to show the effectiveness of combined (multi-modal) embeddings for the purpose of image retrieval and propose an effective solution to leverage style embeddings. Also, our work is the first to study the depth of the mobile counterfeit problem in Google Play Store. We show that it is a common practice than many would think, and often top apps are being targeted. Our analysis shows that the majority of the counterfeits remains in Play Store for long times before they get noticed by the administrators and in some occasions were downloaded over millions of times. }


\section{Dataset}
\label{Sec:Dataset}

We collected our dataset by crawling Google Play Store using a Python crawler between January and March, 2018. The crawler was initially seeded with the web pages of the top free and paid apps as of January, 2018 and it recursively discovered apps by following the links in the seeded pages and the pages of subsequently discovered apps. Such links include apps by the same developer and similar apps as recommended by Google. For each app, we downloaded the metadata such as app name, app category, app description, developer name, and number of downloads as well as the app icon in \emph{.jpg} or \emph{.png} format (of size 300 x 300 x 3 - height, width, and three layers for RGB colour channels). The app icon is the same icon visible in the smartphone once the app is installed and also what users see when browsing Google Play Store. 

We discovered and collected information from {\bf\emph{1,278,297 apps}} during this process. For each app, we also downloaded the app executable in APK format using \emph{Google Play Downloader via Command line}\footnote{https://github.com/matlink/gplaycli} tool by simulating a Google Pixel virtual device. We were able to download APKs for {\bf\emph{1,023,521 apps}} out of the total {\bf\emph{1,278,297 apps}} we discovered. The main reason behind this difference is the \emph{paid apps} for which the APK can't be downloaded without paying. Also, there were some apps that did not  support the virtual device we used. Finally, the APK crawler was operating in a different thread than the main crawler as the APKs download is slower due to their large sizes. As a result, there were some apps that were discovered, yet by the time APK crawler reaches them they were no longer available in Google Play Store. \\ \vspace{-3mm} 

\noindent{\bf \emph{Labelled set}}: To evaluate the performance of various image similarity metrics we require a ground truth dataset that contains similar images to a given image. We used a heuristic approach to shortlist a possible set of visually similar apps and then refined it by manual checking. Our heuristic is based on the fact that there are apps in the app market that have multiple legitimate versions. For example, popular game app \emph{Angry Birds} has multiple versions such as \emph{Angry Birds Rio}, \emph{Angry Birds Seasons}, and \emph{Angry Birds Go}. However, all these apps are using the same characters and as such icons are similar from both content (i.e. birds) and stylistic point of view (i.e. colour and texture). Additionally, they will have similar app descriptions as well.

Thus, we first identified the set of developers who has published more than two apps in app store and one app has at least 500,000 downloads. In the set of apps from the same developer, the app with the highest number of downloads was selected as the \emph{base app}. For each other app in the set, we then calculated the \emph{character level cosine similarity} of their \emph{app name} to the base app name and selected only the apps that had over 0.8 similarity and in the same Google Play Store \emph{app category} as the base app. Through this process we identified 2,689 groups of apps. Finally, we manually inspected each of these groups and checked whether the group consists of actual visually similar apps. In some occasions we found that some groups contained apps that are not visually similar and we discarded those. 

\begin{figure}
\centering
\includegraphics[trim=0cm 0cm 0cm 0cm, clip=true,scale=0.5]{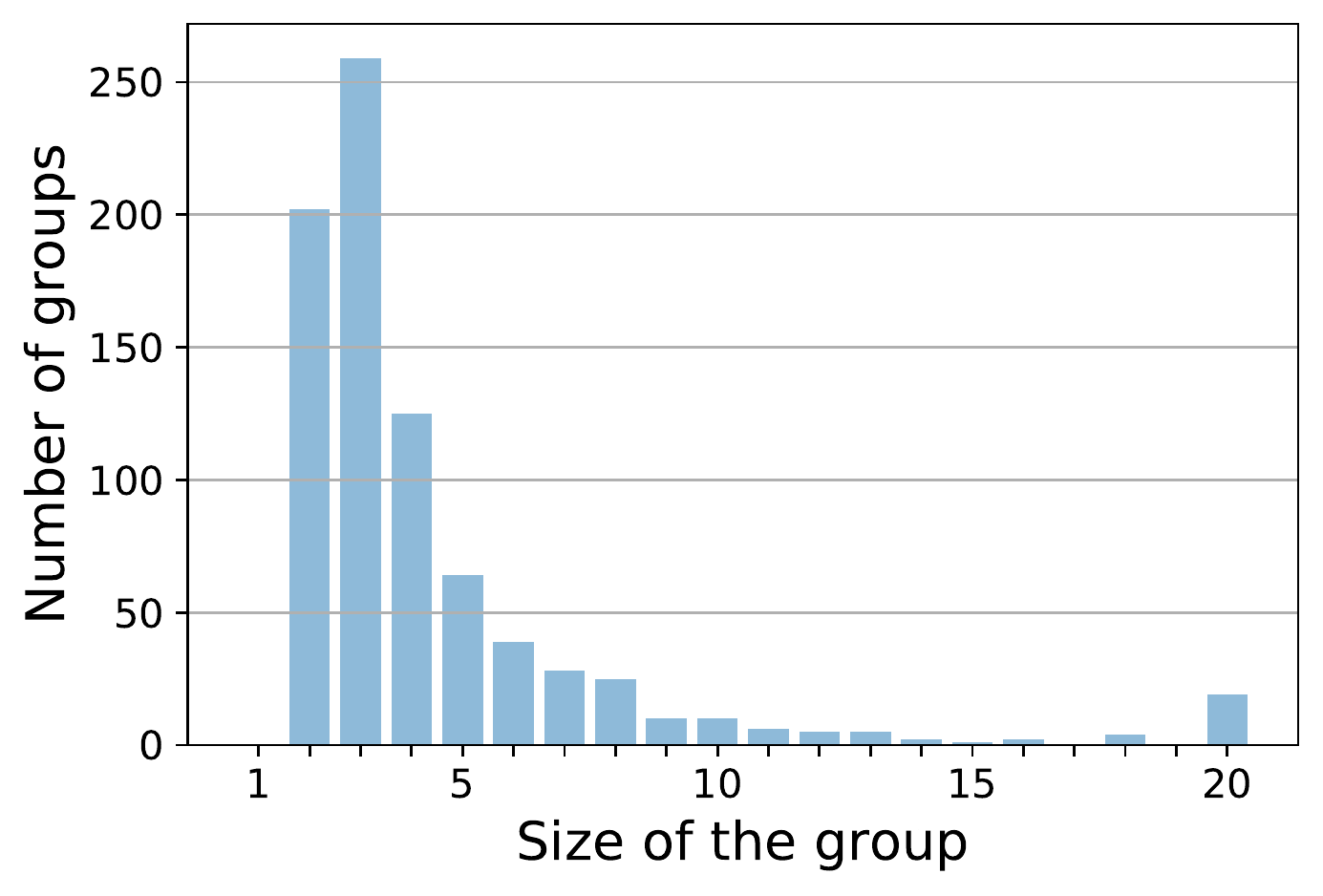}
\caption{Histogram of group sizes in the labelled set}\vspace{-2mm}
\label{Fig:label_dataset_distribution}
\end{figure}

\begin{figure}
\centering
\includegraphics[trim=0cm 0cm 0cm 0cm, clip=true,scale=0.15]{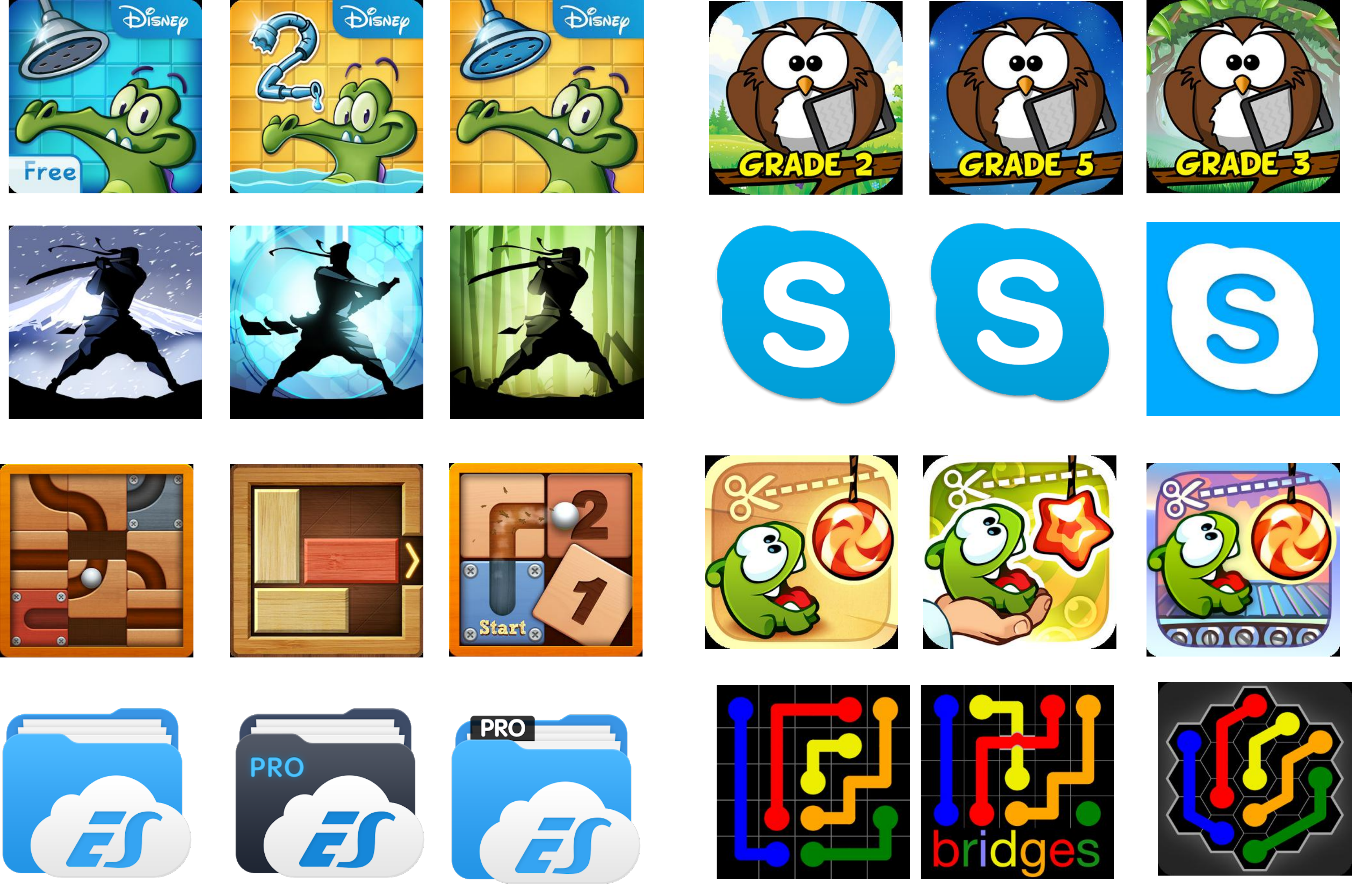}
\caption{Example similar app groups from the labelled set}\vspace{-2mm}
\label{Fig:label_dataset}
\end{figure}

As we describe later, during the evaluation of various image similarity methods, the highest number of neighbours we retrieve for each group is 20. Thus, we ensured that the maximum number of apps in a group was 20 by removing apps from the groups that contained more than 20 apps. To avoid any bias, apps were picked randomly in the removal process.  At the end of this process we had 806 app groups having a total of 3,539 apps as our {\bf \emph{labelled set}}. \textcolor{black}{In  \figurename~\ref{Fig:label_dataset_distribution} we show the histogram of group sizes in the {\bf \emph{labelled set}} and in \figurename~\ref{Fig:label_dataset} we show some example app groups from the {\bf \emph{labelled set}}. The average group size of the labelled set was 4.39}. \\ \vspace{-3mm}

\noindent{\bf \emph{Top-10,000 popular apps}}: To establish a set of potential counterfeits and investigate the depth of app counterfeits problem in Google Play Store, we used top-10,000 apps since counterfeits majorly target popular apps. We selected top-10,000 popular apps by sorting the apps by the number of downloads, number of reviews, and average rating similar to what was proposed in~\cite{seneviratne2015early}. As we describe later in Section~\ref{Sec:Results}, for each app in top-10,000 we retrieved \emph{10-nearest neighbours} by icon and text similarity and investigated those apps for malware inclusion, dangerous permissions, and extra ad library inclusion. \\ \vspace{-3mm}


\noindent{\textcolor{black}{ {\bf \emph{Other image retrieval datasets}}: To benchmark the performance of our proposed combined embeddings approach for measuring image similarity, we use two existing ground-truth datasets; UKBench~\cite{UKBench} and Holidays~\cite{jegou2008hamming}. These two datasets are often used in benchmarking visual image retrieval tasks~\cite{sharif2014cnn,jegou2010aggregating,gong2012iterative}. The specific selection of UKBench and Holidays compared to other image retrieval datasets such as  Oxford5k~\cite{philbin2007object}, Paris6k~\cite{philbin2008lost}, and Sculptures6k~\cite{arandjelovic2011smooth} was based on the fact that these two datasets have a closer average of results per query image to our labelled set (i.e. approximately 3-4 images per query image). Other datasets have much higher values in the range of hundreds.}}

UKBench dataset contains 10,200 images from 2,550 groups. In each group, there are four images taken on the same object from different angles and illumination conditions. In Holidays dataset, there are 1,491 images from 500 groups and each group contains three images on average. The images in each group are taken on a scene with various viewpoints. In both datasets, the first image of each group is taken as the query image to retrieve the nearest neighbours.

\section{Methodology}
\label{Sec:Methodology}

As mentioned before, the main problem we are trying to address is that \emph{``given an app can we find potential counterfeits from a large corpus of apps?''}. Since counterfeit apps focus more on being visually similar to the apps they are trying to impersonate, we mainly focus on finding similar app icons to a given app icon. We also focus on the similarity between text as an additional modality to calculate similarity between apps. We emphasise again that the objective of this work is not to identify app clones or hijacked versions  of apps that will show strong similarities in code level.

\textcolor{black}{Our proposed methodology consists of two high level steps. In the first step, we represent the apps with three types of neural embeddings and their combinations; \textit{content}, \textit{style}, and \textit{text}. In the second step, given an app icon, we retrieve nearest neighbours calculated based on distances between different embeddings with the expectation that if there are any counterfeits to the query app, those will be retrieved in the nearest neighbour search.}

\textcolor{black}{For the rest of this section, we discuss details of the app embedding generation and how we do the nearest neighbour search for various scenarios ({\bf cf.} Section~\ref{SubSec:AppEmbeddings}, Section~\ref{SubSec:Text}, and Section~\ref{SubSec:kNNSearch}). At the end of the section, we describe the baseline methods we use to compare the performance of our proposed method ({\bf cf.} Section~\ref{SubSec:Baselines})}.

\subsection{App Icon Embeddings}
\label{SubSec:AppEmbeddings}

We encode the original app icon image of size $300 \times 300 \times 3$ to a lower dimension for efficient search as well as to avoid false positives happening at large dimensions~\cite{L2_distance_not_good_for_high_diamentions}. We create two types of low dimensional neural representations of the images. \textcolor{black}{For this we use a pre-trained \emph{VGG19}~\cite{vggNet}; a state-of-the-art CNN that is trained on ImageNet dataset~\cite{imagenet_cvpr09}. VGG19 consists of five convolutional blocks followed by two fully connected layers. All convolutional layers use 3x3 kernels as the receptive field and convolution stride is fixed to 1 to preserve spatial resolution after convolution. The numbers of filters used in the five blocks are 64, 128, 256, 512, and 512 respectively. Each block has multiple convolutional layers followed by a max pooling layer which reduces the dimensions of the input. The two fully connected layers consist of 4,096 neurons and followed by the prediction layer with 1,000 classes. All layers use ReLU (Rectified Linear Unit) activations while the prediction layer uses softmax activation.} \\ \vspace{-4mm}



\noindent{\bf \emph{i) Content embeddings}}: To extract the content representations, we fed all 1.2M app icons to the VGG19, and used the \emph{content embeddings}, $C \in \RR^{4096}$, generated at the last fully connected layer of VGG19 (usually called as the ${fc\_7}$ layer) that have shown good results in the past~\cite{fc7_is_good,bell2015siamese}.  \\ \vspace{-3mm} 


\noindent{\bf \emph{ii) Style embeddings}}: As mentioned in Section~\ref{Sec:Introduction}, content similarity itself is not sufficient for counterfeit detection since sometimes developers keep the visual similarity and change the content. For example, if a developer is trying to create a fake app for a popular game that has birds as characters, she can create a game that has the same visual \emph{``look and feel''} and replace birds with cats. Therefore, we require an embedding that represents the style of an image.

Several work demonstrated that the filter responses (also known as feature maps) of convolutional neural networks can be used to represent the style of an image~\cite{gatys2016image,gram}. For example, Gayts et al.~\cite{gatys2016image} used pre-trained convolutional neural networks to transfer the style characteristics of an arbitrary source image to an arbitrary target image. This was done by defining an objective function which captures the \emph{content loss} and the \emph{style loss}. To represent the style of an image, authors used the Gram matrix of the filter responses of the convolution layers. We followed a similar approach and used the fifth convolution layer (specifically $conv5\_1$) of the VGG19 to obtain the style representation of the image, as previous comparable work indicated that $conv5\_1$ provides better performance in classifying artistic styles~\cite{style_classification}. In the process of getting the embeddings for icons, each icon is passed through the VGGNet, and at $conv5\_1$ the icon is convolved with pre-trained filters and activated through $ReLU$ activation function.



More specifically, for an image $\image$, let $\Fl \in\RR^{\Nl\times\Ml}$ be
the filter response of layer $l$, where $\Nl$ denotes the number of filters in layer $l$
and $\Ml$ is the height times width of the feature map. $F^l_{ij}$ is the activation of $i^{th}$ filter at position $j$ in the layer $l$.

Similar to Gayts et al.~\cite{gatys2016image}, to capture style information we use the correlations of the activations calculated by the dot product. That is, for a given image $\image$, let $\Gl\in\RR^{\Nl\times\Nl}$ be the dot product Gram matrix at layer $l$, i.e.

\begin{equation}
   \Glij = \sum_{k=1}^{\Ml}\Flik\Fljk,
  \label{eqn:Glijdotproduct}
\end{equation}
where $\Fl\in\RR^{\Nl\times\Ml}$ is the activation of $I$. Then, $\Gl$ is used as the style representation of an image to retrieve similar images. The $conv5\_1-layer$ of the VGGNet we use has 512 filters, thus the resulting Gram matrix is of size $G^5 \in \RR^{512 \times512}$.




Gram matrix is symmetric as it represents the correlations between the filter outputs. Therefore, we only consider the upper triangular portion and the diagonal of the Gram matrix as our style representation vector, $S \in \RR^{131,328}$.
Though this reduces the dimension of the style vector by about half, the style embedding dimension is much larger compared to the content embeddings, $C \in \RR^{4,096}$. Thus, to further reduce the dimension of style embeddings we used the \emph{very sparse random projection}~\cite{very_sparse_random_projection}. We selected sparse random projection over other dimensionality reduction methods such as PCA and t-SNE due to its computational efficiency~\cite{bingham2001random}. 




More specifically, let $\mathbf{A} \in \RR^{n \times D}$ be our style matrix that contains the style embeddings of a mini batch of $n$ (in the experiments we used $n$=20,000) icons stacked vertically, and $D$ is the dimension of the style vector, which in this case is $131,328$. Then, we create a sparse random matrix $\mathbf{R} \in \RR^{D\times k}$ and multiply it with $A$. 
The elements $r_{ij}$ of $\mathbf{R}$ are drawn according to the below distribution,

\begin{equation} \label{eqn:sparse_matrix}
r_{ij} = \sqrt[4]{D}\left\{\begin{array}{lr}
1, & \text{with prob. } \frac{1}{2\sqrt{D}} \\
 0, & \text{with prob. } 1- \frac{1}{\sqrt{D}} \\
-1, & \text{with prob. } \frac{1}{2\sqrt{D}} \\
 \end{array}\right. 
\end{equation}


At large dimensions, $\mathbf{R}$ becomes sparse as the probability of getting a zero is increasing with $D$. Since sparse matrices are almost orthogonal~\cite{sparse_random_projection_orthogonal,sparse_random_projection_orthogonal2}, multiplication with $\mathbf{R}$, projects $\mathbf{A}$ in another orthogonal dimension.

\begin{equation}
   \mathbf{B} = \frac{1}{\sqrt{k}}\mathbf{A}\mathbf{R} \in \RR^{n \times k}
  \label{eqn:random_production}
\end{equation}

Each row of $\mathbf{B}$ gives a dimensionality reduced vector $S' \in \RR^{k}$ and in our case we used  $k=4,096$ to ensure the size of style embeddings matches the size of the content embeddings. 

\begin{figure*}
\centering
\includegraphics[trim=0.4cm 0cm 0cm 4cm, clip=true, scale=0.4]{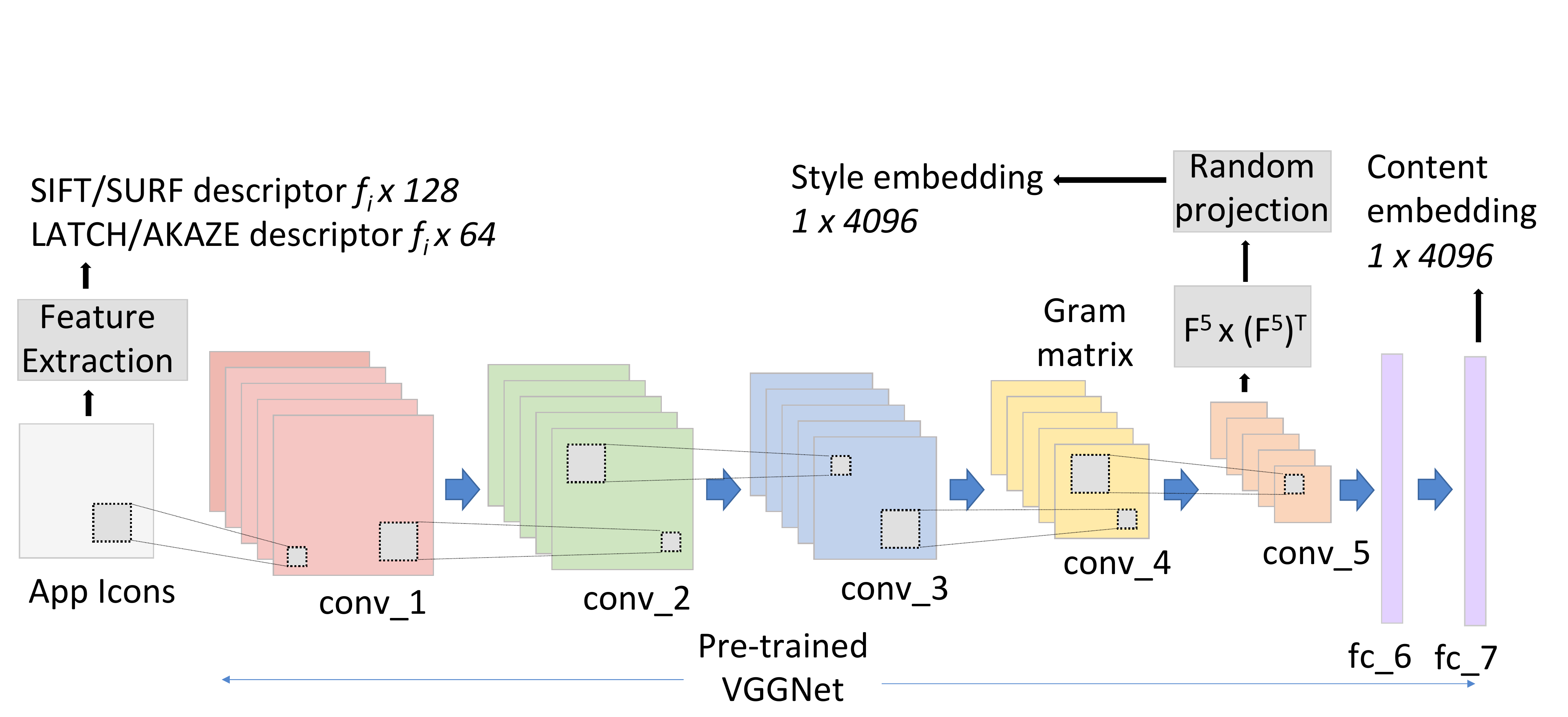}%
\caption{Summary of the image encoding/embeddings generation methodology}
\label{Fig:Encoding}
\end{figure*}





\textcolor{black}{The VGG19 architecture has four layers in the fifth convolutional block. Although we use $conv5\_1$ for main results of the paper, we also did a comparison of the performance of all the four layers in block 5 of VGG19. To check the effect of the pre-trained network architecture, we conducted similar experiments with VGG16 and ResNet50 as well. Additionally, we transform the Gram matrix feature space into different kernel-spaces~\cite{novak2016improving, li2017demystifying} and evaluate the performance. These methods provide extra hyper parameters of the kernel to constrain the metric space. However, we found that almost all these kernel spaces preform similar to the feature space. We present these results in Appendix~\ref{appendix:I} and Appendix~\ref{appendix:II} respectively.}

\subsection{Text Embeddings}
\label{SubSec:Text}

The app descriptions from Google Play Store can contain a maximum of 4000 characters, where the developer describes the apps functionalities and features to the potential users.  As such, counterfeits are likely to show some similarities to original apps' description. To capture this similarity, we first pre-processed the app descriptions using standard text pre-processing methods such as tokenising, converting to lower case, removing stop words, and then trained a Distributed Memory Model of Paragraph Vectors (PV-DM)~\cite{le2014distributed} that created vectors of size 100 to represent each app by training the model over 1,000 epochs. Note that due to the character limitation of 4,000 in text descriptions, we can not create embeddings as large as content and style vectors.



\subsection{Retrieving Similar Apps}
\label{SubSec:kNNSearch}

During the similar app retrieval process we take an app and calculated the required embeddings and search in the encoded space for $k$ \emph{nearest neighbours} using \emph{cosine distance}.\footnote{We also tried the $L_2$ distance in our previous work~\cite{rajasegaran2018neural}. However, it always resulted lower performance compared to the cosine distance.} Let $X_{i}^{y}$ be a vectored representation of an app $i$ using the encoding scheme $y$ (image or text) and $X_{t}^{y}$ be the corresponding representation of the target app we are comparing, we calculate the different distance metrics for different representations as summarised in the first row of Table~\ref{Tab:DistanceMetricSuammry}. Note that both $X_{i}^{y}$ and $X_{t}^{y}$ are vectors of size $n$ (i.e.   $X_{i}^{y}$ = $<x_{1}^{y}, x_{2}^{y}, ..., x_{n}^{y}>$). We used different $k$ values to evaluate our method as we discuss later in Section~\ref{Sec:Results}.



\subsection{Baseline Methods}
\label{SubSec:Baselines}

\textcolor{black}{We compare the performance of our method with several baseline methods. Specifically, we use state-of-the-art image \emph{hashing methods}, \emph{feature-based image retrieval methods}, and SSIM (Structural Similarity). We next describe how each method is used in our setting.} \\ \vspace{-3mm}


\noindent{\bf \emph{i) Hashing methods}}:
Hashing methods we evaluate in this paper include \emph{average}~\cite{AHash}, \emph{difference}~\cite{DHash}, \emph{perceptual}~\cite{PHash}, and \emph{wavelet}~\cite{WHash} hashing. All four methods scale the input app icon into a gray-scale $32 \times 32$ image first and represent the image as a $1024$ dimensional binary feature vector so that it can easily compute the similarity score using hamming distance. Average hash computes a binary vector based on whether pixels' values are greater than the average colour of the image. Difference hash tracks the gradients of the image and perceptual hash evaluates the frequency patterns using discrete cosine transform to generate a binary vector. Wavelet hashing also works in frequency domain, and it calculates a vector using discrete wavelet transformation. \textcolor{black}{Since hashing methods result in binary image representations we use \textit{hamming distance} as the distance metric.} \\ \vspace{-3mm}


\noindent{\bf \emph{ii) Feature-based image retrieval methods}}: 
Feature-based image retrieval methods extract features from an image which are invariant to scale and rotation, and describe them using their neighbouring pixels. Thus, feature-based image retrieval has two steps; \emph{feature detection} and \emph{feature description}. Some algorithms perform both tasks together while others perform them separately. In this paper, we use four feature matching methods; \emph{Scale-Invariant Feature Transform (SIFT)}~\cite{SIFT}, \emph{Speeded-Up Robust Features (SURF)}~\cite{SURF}, \emph{Accelerated KAZE (AKAZE)}~\cite{AKAZE}, and \emph{Learned Arrangements of Three Patch Codes (LATCH)}~\cite{LATCH}. SIFT, SURF, and AKAZE perform both feature detection and description. However, LATCH performs only the feature description task, thus SIFT feature detector is used in LATCH algorithm. SIFT and SURF describe the app icon by a $f_i \times 128$ integer descriptor matrix, where $f_i$ is the number of features detected for app icon $i$. AKAZE and LATCH describe the app icon by a $f_i \times 64$ binary descriptor matrix. After representing all 1.2 million apps by a feature descriptor, feature matching is performed using euclidean distance for integer descriptors (SIFT and SURF) and hamming distance for binary descriptors (AKAZE and LATCH). \\ \vspace{-3mm}

In feature-based methods, the image vector size $n$ depends on the number of features ($f_i^{y}$) detected for app icon $i$ by each method and thus $n$ is not a constant value. Moreover, the distance calculation between the base app $i$ and the target app $t$ is carried out in two steps. First, the closest feature pair for all $f_i^y$ features among $f_t^y$ features is selected using $L2$/hamming distance between the feature descriptors $x_i^y$ and $x_t^y$. Then the total distance is calculated as the sum of the distances between all feature descriptors $x_i^y$ and its closest pair. \\ \vspace{-3mm}

\noindent{\bf \emph{iii) Structural Similarity Index Matrix (SSIM)}}:
SSIM~\cite{SSIM} compares the local pattern of pixel intensities in two images and calculate a similarity score. This method gives a high similarity score even for images with significant pixel-wise difference as it does not compare images point-by-point basis. SSIM does not represent an image by a vector/matrix, thus the whole image is required every comparison. This is one of the drawbacks in SSIM. Therefore, we scale the input app icons into a gray-scale $32 \times 32$ images and calculate the similarity score using SSIM. 

\textcolor{black}{In \figurename~\ref{Fig:Encoding}, we show an overview of our icon encoding process and in Table~\ref{Tab:DistanceMetricSuammry} we show a summary of all the different embeddings/encoding methods and the distance metrics we used including the baselines.}




\begin{table}

\centering
\caption{Summary of different encoding methods and corresponding distance metrics} \vspace{-3mm}
\begin{tabular}{p{2.1cm}p{1.5cm}p{4.1cm}}

\specialrule{.12em}{1em}{0em}
 \textbf{Encoding Method} & \textbf{Size (n)} & {\textbf{Distance function}} \\ \hline
 \specialrule{.12em}{0em}{0em}
 
 \rowcolor{gray1}
\multicolumn{3}{l}{\textbf{Neural embeddings (Cosine distance)}} \\ \hline 
\specialrule{.12em}{0em}{0em}
\small Content ($C_{cos}$)  &  \small 4,096 &  $1 - \frac{X_{i}^{cont}.X_{t}^{cont}}{ \norm{X_{i}^{cont}}_2 \norm{X_{t}^{cont}}_2}$  \\ 
\small Style ($S_{cos}$)&   \small 4,096 &    $1 - \frac{X_{i}^{style}.X_{t}^{style}}{ \norm{X_{i}^{style}}_2 \norm{X_{t}^{style}}_2}$ \\ 
\small Text ($T_{cos}$)& \small 100 &  $1 - \frac{X_{i}^{text}.X_{t}^{text}}{ \norm{X_{i}^{text}}_2 \norm{X_{t}^{text}}_2}$ \\ 
\small $\alpha\;$Content+$\beta\;$Style&  \small 8,192 & $\alpha\;C_{cos}+\beta\;S_{cos}$ \\ 
\small $\alpha\;$Content+$\beta\;$Style&  \small 8,292 & $\alpha\;C_{cos}+\beta\;S_{cos} + \gamma\;T_{cos}$  \\ 
 \small $\qquad$+$\gamma\;$Text  & & \\
\specialrule{.12em}{0em}{0em}

\rowcolor{gray1}
\multicolumn{3}{l}{\textbf{Hashing methods (Hamming distance)}} \\ \hline 
\specialrule{.12em}{0em}{0em}
\small Average & \small 1,024 &  \footnotesize $ \norm{X_{i}^{avg} \oplus X_{t}^{avg}}_1 $ \\ 
\small Difference &  \small 1,024 &  \footnotesize $ \norm{X_{i}^{diff} \oplus X_{t}^{diff}}_1 $  \\ 
\small Perceptual & \small 1,024 &  \footnotesize $ \norm{X_{i}^{perc} \oplus X_{t}^{perc}}_1 $ \\ 
\small Wavelet & \small 1,024 &  \footnotesize $ \norm{X_{i}^{wave} \oplus X_{t}^{wave}}_1 $  \\ 
\specialrule{.12em}{0em}{0em}

\rowcolor{gray1}
\multicolumn{3}{l}{\textbf{Feature based methods ($L_2$ distance)}} \\ \hline
\specialrule{.12em}{0em}{0em}
\small SIFT &\footnotesize  $f_i^{sift}\times 128$& \footnotesize $ \sum\limits_{\substack{x_{i} \in X_{i}}} \min\limits_{\substack{x_{t} \in X_{t}}}[\norm{x_{i}^{sift} - x_{t}^{sift}}_2 ]$ \\ 
\small SURF &  \footnotesize $f_i^{surf}\times 128$ & \footnotesize $ \sum\limits_{\substack{x_{i} \in X_{i}}} \min\limits_{\substack{x_{t} \in X_{t}}}[\norm{x_{i}^{surf} - x_{t}^{surf}}_2 ] $ \\ 
\specialrule{.12em}{0em}{0em}
\rowcolor{gray1}
\multicolumn{3}{l}{\textbf{Feature based methods (Hamming distance)}} \\ \hline 
\specialrule{.12em}{0em}{0em}
\small AKAZE & \footnotesize $f_i^{akaze}\times 64 $ & \footnotesize $ \sum\limits_{\substack{x_{i} \in X_{i}}} \min\limits_{\substack{x_{t} \in X_{t}}}[\norm{x_{i}^{akaze} \oplus  x_{t}^{akaze}}_2 ]$ \\ 
\small LATCH & \footnotesize $f_i^{latch}\times 64 $& \footnotesize $ \sum\limits_{\substack{x_{i} \in X_{i}}} \min\limits_{\substack{x_{t} \in X_{t}}}[\norm{x_{i}^{latch} \oplus  x_{t}^{latch}}_2 ]$  \\ 
\specialrule{.12em}{0em}{0em}

\rowcolor{gray1}
\multicolumn{3}{l}{\textbf{Structural similarity}} \\ \hline 
\specialrule{.12em}{0em}{0em}
\small SSIM &  \multicolumn{2}{p{5.6cm}}{\small Directly returns a dissimilarity value between 0 and 1} \\ \hline
\specialrule{.12em}{0em}{0em}

\end{tabular}
\label{Tab:DistanceMetricSuammry}
\end{table}

\section{Results}
\label{Sec:Results}

\subsection{Evaluation of Embeddings}

To quantify the performance of the different encodings and the proposed multi-modal embeddings, we evaluate them in four different test scenarios using multiple datasets. In each scenario, for a given query encoding/embedding, we retrieved $k$-\emph{nearest neighbours} (k-NN)  based on the distances considered in Table~\ref{Tab:DistanceMetricSuammry}. We tested four values of $k$; 5, 10, 15, and 20. The four scenarios are:  \\ \vspace{-1mm}



\noindent{\bf{i) Holidays dataset: }} As mentioned before, the Holidays dataset contains 1,491 images from 500 groups. We took the encoded representation of the first image from each group as the query to search the entire corpus and retrieved the \emph{k-nearest neighbours}. \\ \vspace{-3mm}


\noindent{\bf{ii) UKBench dataset: }} UKBench dataset contains 10,200 images from 2,550 groups. The encoding of the first image in each group was taken as the query to retrieve the \emph{k-nearest neighbours} from the entire dataset. \\  \vspace{-3mm}

 \begin{table*}  \centering 
\footnotesize{
    \begin{tabular}{p{0.1cm}p{0.9cm}p{0.8cm}p{0.8cm}p{0.8cm}p{0.8cm}p{0.8cm}p{0.8cm}p{0.8cm}p{0.8cm}p{0.8cm}p{0.8cm}p{0.8cm}p{0.8cm}p{0.8cm}}
        & & \rot{\bf Average} & \rot{\bf Difference} & \rot{\bf Perceptual} & \rot{\bf Wavelet} 
        & \rot{\bf SIFT} & \rot{\bf SURF} & \rot{\bf LATCH} 
        & \rot{\bf AKAZE} & \rot{\bf SSIM} & \rot{$\bm{C_{cos}}$} & \rot{$\bm{S_{cos}}$} & \rot{$\bm{C_{cos}+\beta S_{cos}}$} & \rot{\shortstack[l]{\bm{$C_{cos}+\beta S_{cos}$}\\$\bm{+\gamma T_{cos}}$}} \\
        \cmidrule[1pt]{2-15}
        & \bf{5-NN}             & 24.56 & 22.68 & 21.60 & 24.48  &  33.00 & 31.12 & 29.16 & 31.00& 21.88  & 46.36 & 46.72 & \textbf{47.92} & N/A \\
        & \bf{10-NN}               & 13.08 & 11.74 & 10.96 &12.98 &  17.58 & 16.66 & 15.18 & 15.90 & 11.26  & 25.28 & 25.24 & \textbf{25.92} & N/A \\

        & \bf{15-NN}             & 9.00 & 8.08 & 7.36 & 8.92 &  12.15 & 11.55 & 10.43 & 10.91 & 7.76 & 17.47 & 17.25 & \textbf{17.89} & N/A \\
                 \rott{\rlap{\bf Holidays}}
        & \bf{20-NN} & 6.95 & 6.19 & 5.58 & 6.83 &  9.34 & 8.91 & 7.99 & 8.31 & 5.98  & 13.31 & 13.13 & \textbf{13.57} & N/A  \\

        \cmidrule{2-15}
        
                & \bf{5-NN}             & 27.29 & 22.44 & 21.63 & 26.37 &  55.27 & 52.97 & 44.82 & 41.77 & 28.46  & \textbf{70.22 }& 65.02 & 70.06 & N/A  \\
        & \bf{10-NN}               & 15.01 & 11.70 & 10.97 & 14.26 &  28.99 & 27.93 & 23.72 & 21.98 & 15.22  & \textbf{36.90} & 33.86 & 36.62 & N/A  \\

        & \bf{15-NN}              & 10.51 & 7.95 & 7.38 & 9.99  &  19.82 & 19.12 & 16.22 & 15.04 & 10.55  & \textbf{25.03} & 22.95 & 24.79 & N/A  \\
                 \rott{\rlap{\bf UKBench}}
        & \bf{20-NN} & 8.18 & 6.08 & 5.59 & 7.75 &  15.11 & 14.60 & 12.37 & 11.47 & 8.17  & \textbf{18.97} & 17.40 & 18.75 & N/A  \\

        \cmidrule{2-15}
        
        & \bf{5-NN}           & 45.14 & 48.41 & 47.62 & 44.44  &  48.92 & 47.67 & 46.63 & 44.22 & 45.34  & 56.43 & 60.57 & 62.23 & \textbf{64.76}  \\
        & \bf{10-NN}               & 23.98 & 28.10 & 27.42 & 25.50 &  26.79 & 27.05 & 26.34 & 25.07 & 25.59  & 33.69 & 35.84 & 36.04 & \textbf{38.47}  \\

        & \bf{15-NN}              & 18.59 & 19.92 & 19.45 & 18.08  &  18.86 & 19.00 & 18.45 & 17.54 & 18.06  & 24.05 & 25.45 & 25.57 & \textbf{27.19}  \\
                 \rott{\rlap{\bf Labelled}}
        & \bf{20-NN} & 14.52 & 15.56 & 15.24 & 14.16  &  14.57 & 14.69 & 14.24 & 13.5 & 14.09  & 18.69 & 19.75 & 19.86 & \textbf{21.09}  \\

        \cmidrule{2-15}
        
                & \bf{5-NN}            & 34.89 & 38.01 & 37.07 & 34.17  &  38.23 & 39.13 & 37.32 & 36.87 & 37.39  & 45.66 & 50.67 & 50.91 & \textbf{55.96}  \\
        & \bf{10-NN}               & 19.43 & 21.53 & 20.79 & 19.08 &  21.82 & 22.10 & 21.09 & 20.81 & 20.73  & 26.08 & 29.65 & 29.81 & \textbf{32.99}  \\
\rott{\rlap{\bf All}}
        & \bf{15-NN}              & 13.69 & 15.30 & 14.74 & 13.32  &  15.31 & 15.52 & 14.82 & 14.63 & 14.47  & 18.35 & 21.00 & 21.12 & \textbf{23.46}  \\
                 
        & \bf{20-NN} & 10.63 & 11.89 & 11.40 & 10.36  &  11.87 & 11.97 & 11.46 & 11.33 & 11.15  & 14.18 & 16.15 & 16.31 & \textbf{18.23 } \\

        \cmidrule[1pt]{2-15}

    \end{tabular}
    \caption{precision@k for all test scenarios (NN* - Nearest Neighbours)} 
    \label{Tab:Precision}}
    \vspace{-5mm}
\end{table*}

\begin{table*} \centering
\footnotesize
    \begin{tabular}{p{0.1cm}p{0.9cm}p{0.8cm}p{0.8cm}p{0.8cm}p{0.8cm}p{0.8cm}p{0.8cm}p{0.8cm}p{0.8cm}p{0.8cm}p{0.8cm}p{0.8cm}p{0.8cm}p{0.8cm}}
        & & \rot{\bf Average} & \rot{\bf Difference} & \rot{\bf Perceptual} & \rot{\bf Wavelet} 
        & \rot{\bf SIFT} & \rot{\bf SURF} & \rot{\bf LATCH} 
        & \rot{\bf AKAZE} & \rot{\bf SSIM} & \rot{$\bm{C_{cos}}$} & \rot{$\bm{S_{cos}}$} & \rot{$\bm{C_{cos}+\beta S_{cos}}$} & \rot{\shortstack[l]{\bm{$C_{cos}+\beta S_{cos}}$\\$\bm{+\gamma T_{cos}}$}} \\
        \cmidrule[1pt]{2-15}
        & \bf{5-NN}             &41.18 & 38.03 & 36.22  & 41.05  &  55.33&  52.18 & 48.89 & 51.98 & 36.69  & 77.73 & 78.34 & \textbf{80.35} & N/A \\
        & \bf{10-NN}               & 43.86 & 39.37 & 36.75  & 43.53  &  58.95 &  55.87 & 50.91 & 53.32 & 37.76  & 84.78 & 84.64 & \textbf{86.92} & N/A \\

        & \bf{15-NN}             & 45.27 & 40.64 & 37.02  & 44.87  &  61.10 &  58.08 & 52.45 & 54.86 & 39.03  & 87.86 & 86.79 & \textbf{90.01} & N/A \\
                 \rott{\rlap{\bf Holidays}}
        & \bf{20-NN} & 46.61 & 41.52 & 37.42  & 45.81  &  62.64 &  59.76 & 53.59 & 55.73 & 40.11  & 89.27 & 88.06 & \textbf{91.01} & N/A  \\

        \cmidrule{2-15}
        
                & \bf{5-NN}             & 34.11 & 28.05 & 27.04 & 32.96  &  69.09 & 66.22 & 56.03 & 52.23 & 35.58 & \textbf{87.78} & 81.27 & 87.58 & N/A  \\
        & \bf{10-NN}               & 37.51 & 29.25 & 27.42 & 35.66 &  72.47 & 69.84 & 59.3 & 54.96 & 38.03  & \textbf{92.25} & 84.65 & 91.54 & N/A  \\

        & \bf{15-NN}              & 39.41 & 29.82 & 27.69 & 37.46 &  74.34 & 71.69 & 60.83 & 56.4 & 39.56  & \textbf{93.85} & 86.08 & 92.96 & N/A  \\
                 \rott{\rlap{\bf UKBench}}
        & \bf{20-NN} & 40.92 & 30.82 & 27.93 & 38.73  &  75.57 & 72.99 & 61.83 & 57.34 & 40.86  & \textbf{94.83} & 86.98 & 93.75 & N/A  \\

        \cmidrule{2-15}
        
        & \bf{5-NN}           & 51.40 & 55.35 & 54.22 & 50.61  &  55.43 & 54.28 & 53.09 & 50.35 & 51.60 & 64.26 & 68.97 & 69.82 & \textbf{73.75}  \\
        & \bf{10-NN}               &59.17 & 64.08 & 62.44 & 58.07 &  61.00 & 61.60 & 59.99 & 57.11 & 58.24  & 76.72 & 81.63 & 82.09 & \textbf{87.62}  \\

        & \bf{15-NN}              & 63.49 & 68.04 & 66.46 & 61.77  &  64.42 & 64.91 & 63.04 & 59.93 & 61.66  & 82.17 & 86.95 & 87.34 & \textbf{92.88}  \\
                 \rott{\rlap{\bf Labelled}}
        & \bf{20-NN} & 66.12 & 70.90 & 69.40 & 64.48  &  66.37 & 66.91 & 64.88 & 61.51 & 64.12  & 85.14 & 89.94 & 90.45 & \textbf{96.07 } \\

        \cmidrule{2-15}
        
                & \bf{5-NN}            & 39.73 & 42.29 & 42.22 & 38.91  &  43.29 & 44.30 & 42.47 & 41.96 & 42.55  & 51.99 & 57.70 & 57.98 & \textbf{63.72}  \\
        & \bf{10-NN}               & 44.25 & 49.03 & 47.36 & 43.46  &  49.42 & 50.04 & 48.01 & 47.36 & 47.19  & 59.40 & 67.53 & 67.90 & \textbf{75.13}  \\
\rott{\rlap{\bf All}}
        & \bf{15-NN}              & 46.76 & 52.27 & 50.35 & 45.49  &  52.02 & 52.73 & 50.61 & 49.93 & 49.39  & 62.70 & 71.74 & 72.14 & \textbf{80.16}  \\
                 
        & \bf{20-NN} & 48.43 & 54.14 & 51.94 & 47.19  &  53.75 & 54.23 & 52.16 & 51.57 & 50.75  & 64.59 & 73.58 & 74.29 & \textbf{83.05}  \\

        \cmidrule[1pt]{2-15}

    \end{tabular}
    \caption{recall@k for all test scenarios (NN* - Nearest Neighbours)}
     \label{Tab:Recall}
\end{table*}

\noindent{\bf{iii) Apps - Labelled set only: }} Our labelled set contains 3,539 images from 806 groups. From each group, the \emph{base app icon embedding} ({\bf{cf.}} Section~\ref{Sec:Dataset}) was taken as the query to retrieve the  \emph{k-nearest} icons by searching through the remaining 3,538 icons.  \\  \vspace{-3mm}
 
\noindent{\bf{iv) Apps -  Labelled set and all remaining icons and text: }} This dataset contains 1.2M images including the images in the labelled set. The embedding of the base app icon of each group in the labelled set was taken as the query to search the entire image set and retrieve the \emph{k-nearest neighbours}. For the last distance metric that contained text, we used the text embeddings generated for all 1.2M app descriptions using the method described in Section~\ref{SubSec:Text}, in addition to the image based content and style embeddings.\\  \vspace{-1mm}

\normalsize

The intuition behind above test scenarios is that if the encoding/embedding is a good representation of the image (or text), the \emph{k-nearest neighbours} we retrieve must be from the same group as the query image (or text). Thus, for each scenario, we present {\bf\emph{precision@k}} and {\bf\emph{recall@k}}, where $k\in\{5,10,15,20\}$, as the performance metrics. {\bf\emph{Precision@k}} gives the percentage of relevant images among the retrieved images as defined in (\ref{eqn:Precision}). {\bf\emph{Recall@k}} is the percentage of relevant images that have been retrieved out of the all relevant images as defined in (\ref{eqn:Recall}). For the last distance metric in Table~\ref{Tab:DistanceMetricSuammry}, we considered the base app itself as the querying item (represented by both image and text embeddings).

\footnotesize{
\begin{equation}\label{eqn:Precision}
precision@k = \frac{\mid\{relevant\;images\} \cap \{retrieved \; images\} \mid}{\mid\{retrieved\;  images\} \mid}*100\%
\end{equation}
\begin{equation}\label{eqn:Recall}
recall@k = \frac{\mid\{relevant\; images\} \cap \{retrieved \; images\} \mid}{\mid\{ relevant\;  images\} \mid}*100\%
\end{equation}
}

\normalsize

We present precision@k and recall@k values for all four test scenarios in Table~\ref{Tab:Precision} and Table~\ref{Tab:Recall}, respectively. To choose the best $\beta$ and $\gamma$ values in multi-modals neural embeddings, we varied $\beta$ and $\gamma$ from 1 to 10 with an interval of one. We achieved the best results when $\beta=5$ and $\gamma=4$ and we report those results in Tables~\ref{Tab:Precision} and \ref{Tab:Recall}. The main takeaway messages from results in these two tables can be summarised as below.

\begin{itemize}
\item In all four datasets, neural embedding methods outperform hashing and feature-based methods. For example, for all four k-NN scenarios, the style embeddings have approximately 4\%--14\% and 11\%--26\% higher performance in precision@k and recall@k in all apps dataset compared to  hashing and feature-based baseline methods. 

\item In UKBench and Holidays datasets, content, style, and combined embeddings increase \emph{precision@k} and \emph{recall@k} by 10\%--15\% and 12\%--25\%, respectively when retrieving five nearest neighbours. Combining style embeddings with content embeddings achieves 12\% higher precision@k and 14\% higher recall@k in all app dataset compared to hashing and feature-based baselines when $k=5$. Only scenario where combined content and style embeddings did not outperform all other methods is the UKBench dataset. A possible reason for this is that UKBench dataset contains images that are similar to the ImageNet dataset used to pre-train the VGGNet.

\item It is also noticeable that adding text embedding further increases the performance by 3\%-5\% and 6\%-7\% in terms of precision@k and recall@k, respectively, compared to the best neural embedding method when $k\in\{5,10\}$. This method is not applicable for UKBench and Holidays datasets as there are no associated  text descriptions for those images. 

\item Results also show that increasing the k value in top-k scenarios increases the recall@k, however, significantly decreases precision@k. The main reason is that average number of images per groups in all four datasets is less than 5 and thus the number of false positive images in the retrieved image set increases when k increases. 
\end{itemize}

\begin{figure}
\centering
\includegraphics[scale = 0.2]{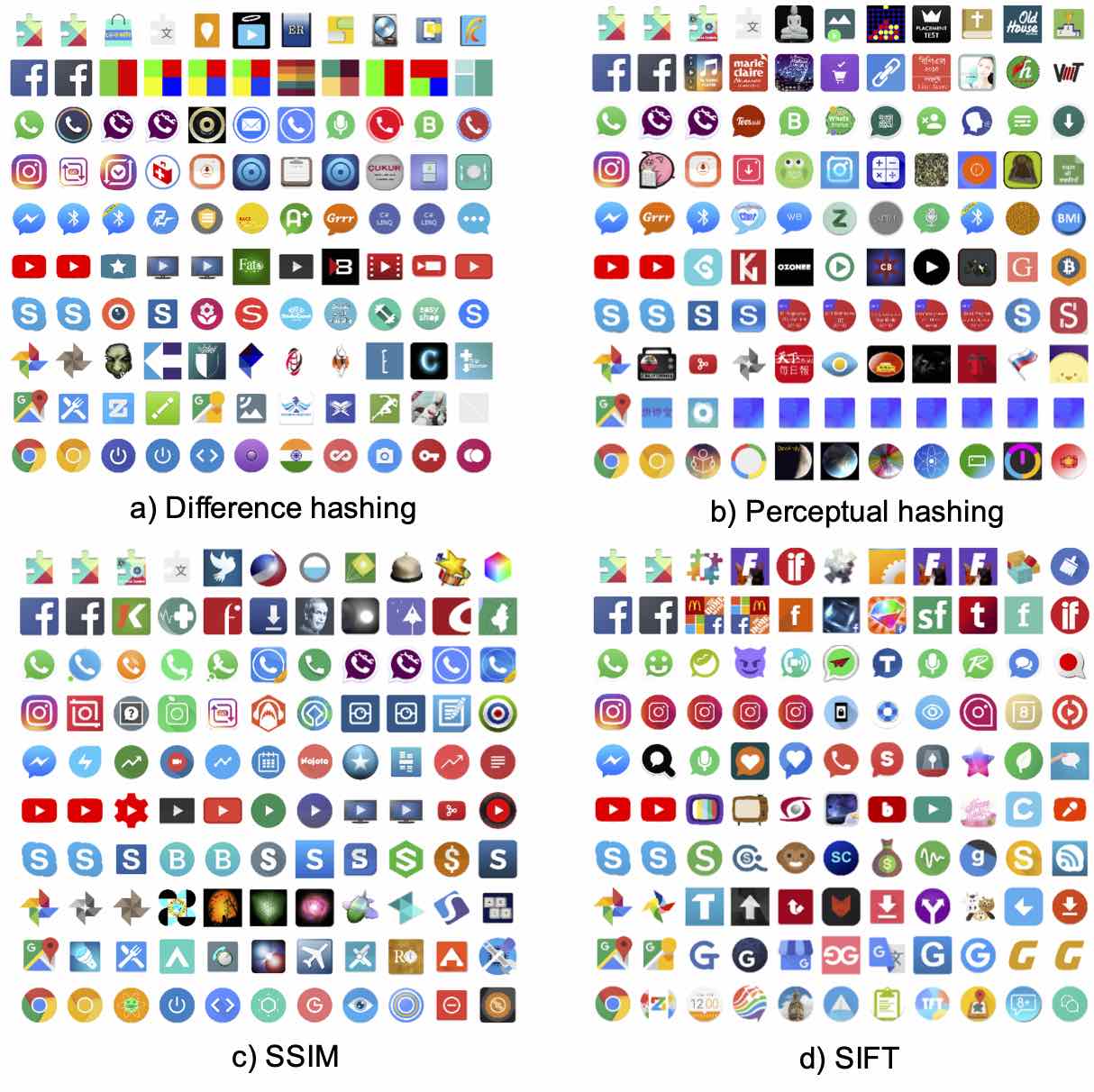}%
\caption{10-Nearest neighbours of the top-10 popular apps (hashing, feature-based, and structural similarity methods)}
\label{Fig:top_10_embeddings_classical}
\end{figure}

\begin{figure}
\centering
\includegraphics[scale = 0.2]{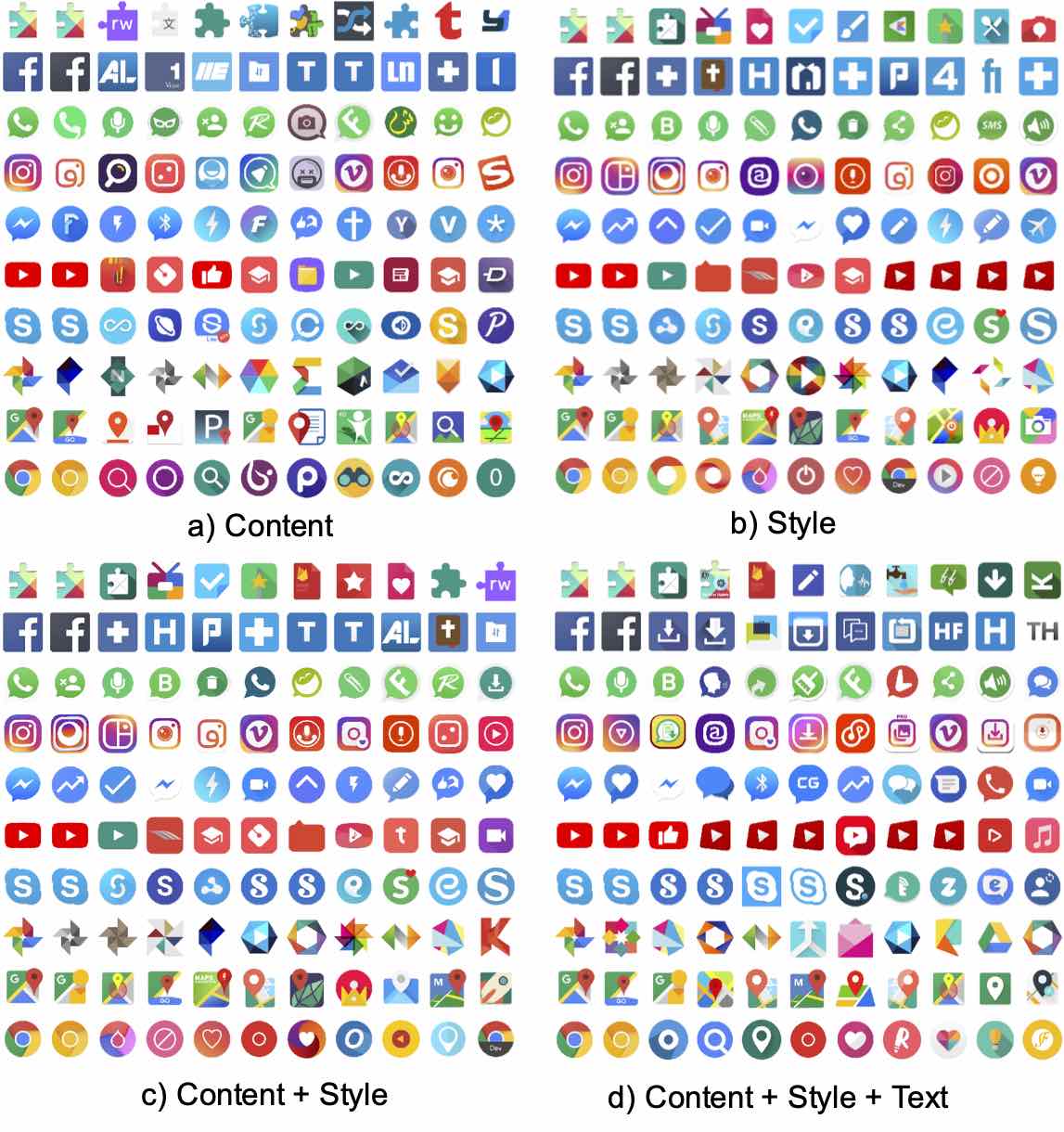}%
\caption{10-Nearest neighbours of the top-10 popular apps (neural embeddings)}
\label{Fig:top_10_embeddings}
\end{figure}

To elaborate further on the performance of the embeddings qualitatively, in \figurename~\ref{Fig:top_10_embeddings_classical} and \figurename~\ref{Fig:top_10_embeddings}, we present the 10-nearest neighbours we retrieved using difference hashing, feature-based methods, and different neural embeddings for the top-10 most popular apps in Google Play Store. In most of the cases, the methods presented in \figurename~\ref{Fig:top_10_embeddings_classical} do not identify visually similar apps apart from the first 1-2 similar apps (E.g. row 9 - Google Maps). We can also observe that if the orientations of the two images are similar, hashing and SSIM methods are able to retrieve the images (E.g. row 3 - Whats app and row 7 - Skype). Neural embeddings based methods in \figurename~\ref{Fig:top_10_embeddings} have identified better fits in several cases (E.g. row 1 - Google Play Services and row 9 - Google Maps). 

In particular, in \figurename~\ref{Fig:top_10_embeddings}-(b) style embeddings have retrieved app icons that have the same \emph{``look and feel''} in terms of colour (e.g. Facebook Messenger in row 5). The improvement provided by combining text embeddings is visible in some of the cases in \figurename~\ref{Fig:top_10_embeddings}-(d). For instance, Skype in row 7 and YouTube in row 6.

\textcolor{black}{We also experimented the performance of style embeddings generated from other convolutional layers of VGG19 and layers from VGG16, and ResNet50. We used layers in the fifth convolution block of each architecture for the experiment as deeper layers have proven to perform better in representing the style of an image. We found out that although there are variations among different individual layers, all three architectures perform in a similar manner. For clarity we discuss these results in Appendix~\ref{appendix:I}. Additionally, we also experimented with projecting style embeddings into different kernel spaces. This allows us to model the different characteristics of the metric space, which cannot be captured in the feature space. Subsequently, we project the feature space embeddings into Squared exponential kernel, and polynomial kernel, and we also shift the features by a fixed bias to move the embeddings in the feature space. However, there is no significant gain in the model performance with kernel space embeddings. There results are presented in Appendix~\ref{appendix:II}.}

\subsection{Retrieving Potential Counterfeits}

We next use the embeddings that performed best (Content$_{cos}$ + $\beta$Style$_{cos}$ + $\gamma$Text$_{cos}$ where $\beta=5$ and $\gamma=4$) to retrieve similar apps for {\bf \emph{top-10,000 popular apps}}  and check the availability of malware, as spreading malware can be one of the main objectives behind publishing counterfeit apps. In this analysis, we focus only on the popular apps since they usually are the main targets of counterfeits. For each app in the {\bf \emph{top-10,000 popular apps}} we retrieved 10-nearest neighbours apps (in terms of both visual and text similarity) from the corpus of 1.2 million apps that are not from the same developer. 

\begin{figure}
\centering
\includegraphics[scale=0.5]{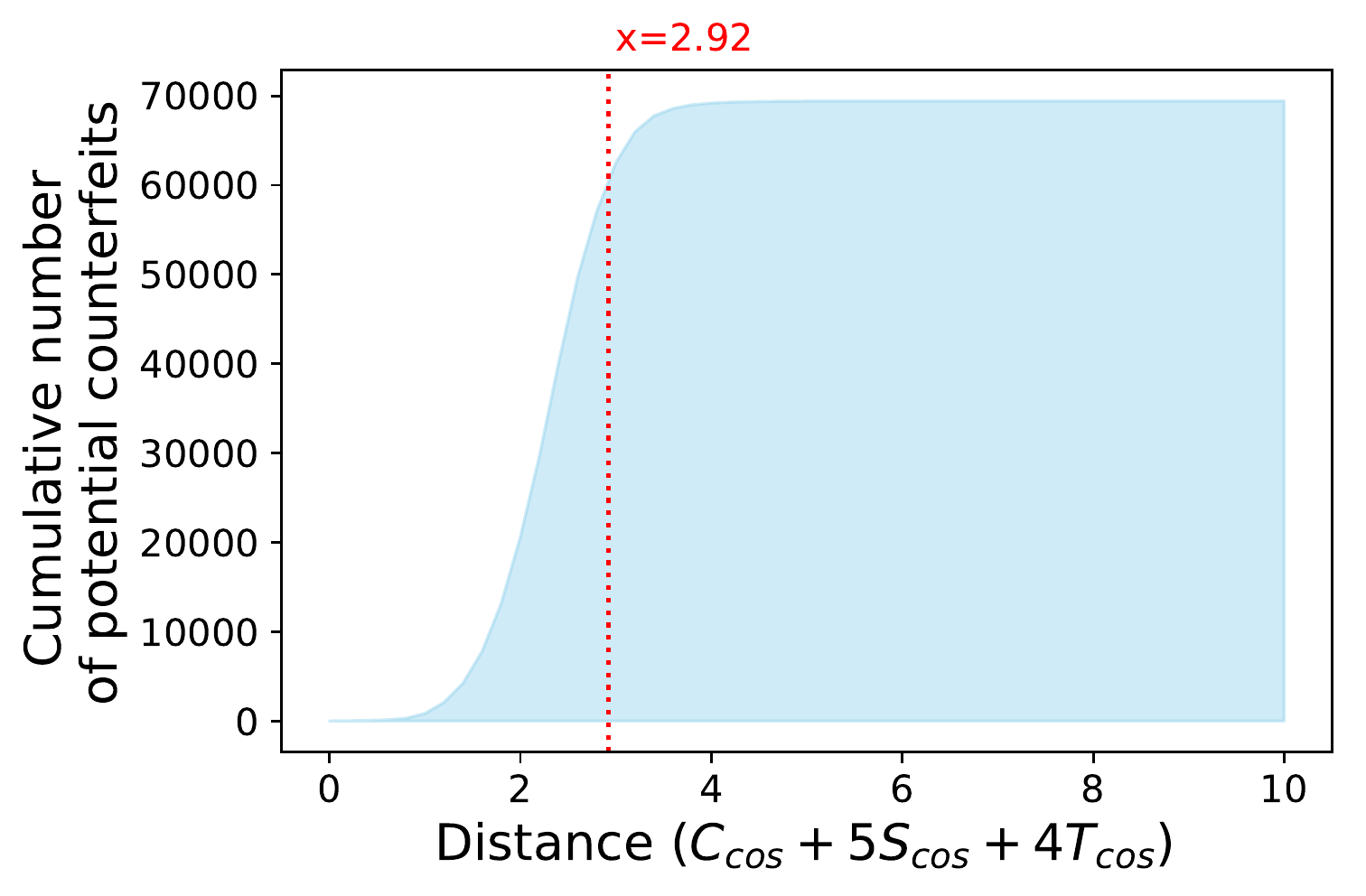}%
\caption{Cumulative number of apps against the multi-modal embeddings based distance}
\label{Fig:Distance}
\end{figure}

However, the 10-nearest neighbour search is forced to return $10$ nearest apps, irrespective of the distance. As such, there can be cases where the nearest neighbour search returns apps that are very far from the query app. Thus, we applied a distance threshold to further narrow down the retrieved results. From the retrieved 10 results for each query app, we discarded the results that are having  distances greater than a empirically decided threshold. The threshold was chosen as the knee-point~\cite{satopaa2011finding} of the cumulative distribution of all the distances with the original apps as shown in Fig.~\ref{Fig:Distance}. The exact value of threshold is 2.92. Note that the maximum distance that can occur for the embedding we consider;  Content$_{cos}$ + $\beta$Style$_{cos}$ + $\gamma$Text$_{cos}$ is 10 since the maximum value of cosine distance is one. \textcolor{black}{It is possible to further lower this distance threshold and have more conservative retrievals with high precision or increase this threshold and  reach a high recall. This is a decision at the hand of the app market operators depending on the effort they want to put into further scrutinise the possible counterfeits.}



\begin{figure}
\centering
\includegraphics[scale=0.03]{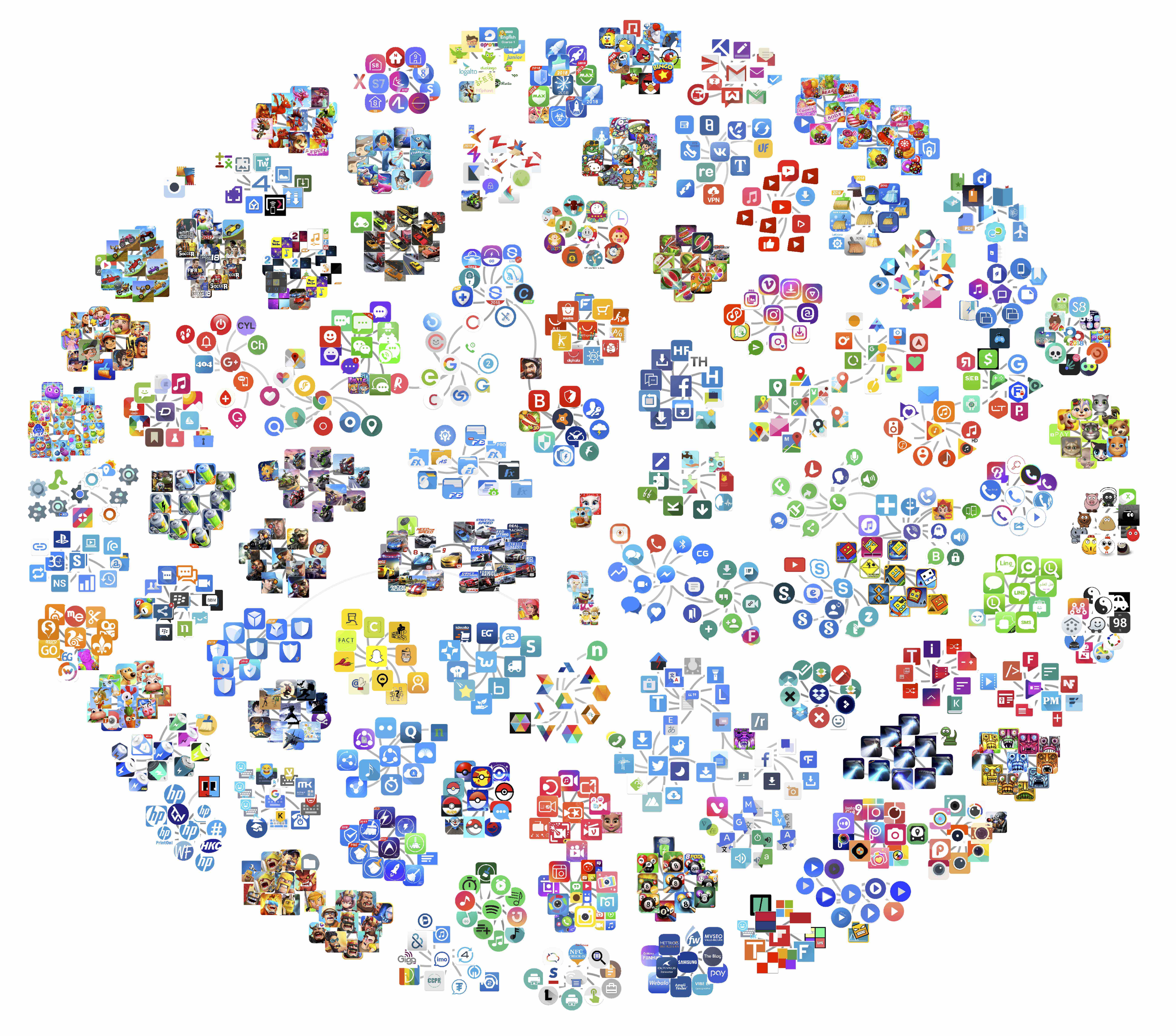}%
\caption{Graph-based visualisation of top-100 apps and similar app icons (small clusters in this figure do not contain apps from the same developer)}
\label{Fig:graph}
\end{figure}

This process returned 60,638 unique apps that are potentially counterfeits of one or more apps with in top-10,000 popular apps. Out of this 60,638 we had APK files for 49,608 apps. In \figurename~\ref{Fig:graph}, we show a graph-based visualisation of the app icons of potential counterfeits we identified for top-100 popular apps. The centre node of each small cluster represent an app in top-100 and the connected apps to that are the similar apps we identified for that particular app. As the figure shows, many of the similar apps retrieved show high visual similarity to the original app. 

\subsection{Malware Analysis}

We then checked each of the 49,608 potential counterfeits using the private API of the online malware analysis tool \emph{VirusTotal}.\footnote{https://www.virustotal.com} VirusTotal scans the APKs with over 60 commercial anti-virus tools (AV-tools) in the likes of AVG, Avast, Microsoft, BitDefender, Kaspersky, and  McAfee and provides a report on how many of those tools identified whether the submitted APKs contain malware.  We used the private API of VirusTotal because there is a rate (4 requests per minute) and size (32 MB) limitations of the number of binaries that can be analysed using the public API.

\begin{figure}
\centering
\includegraphics[scale=0.5]{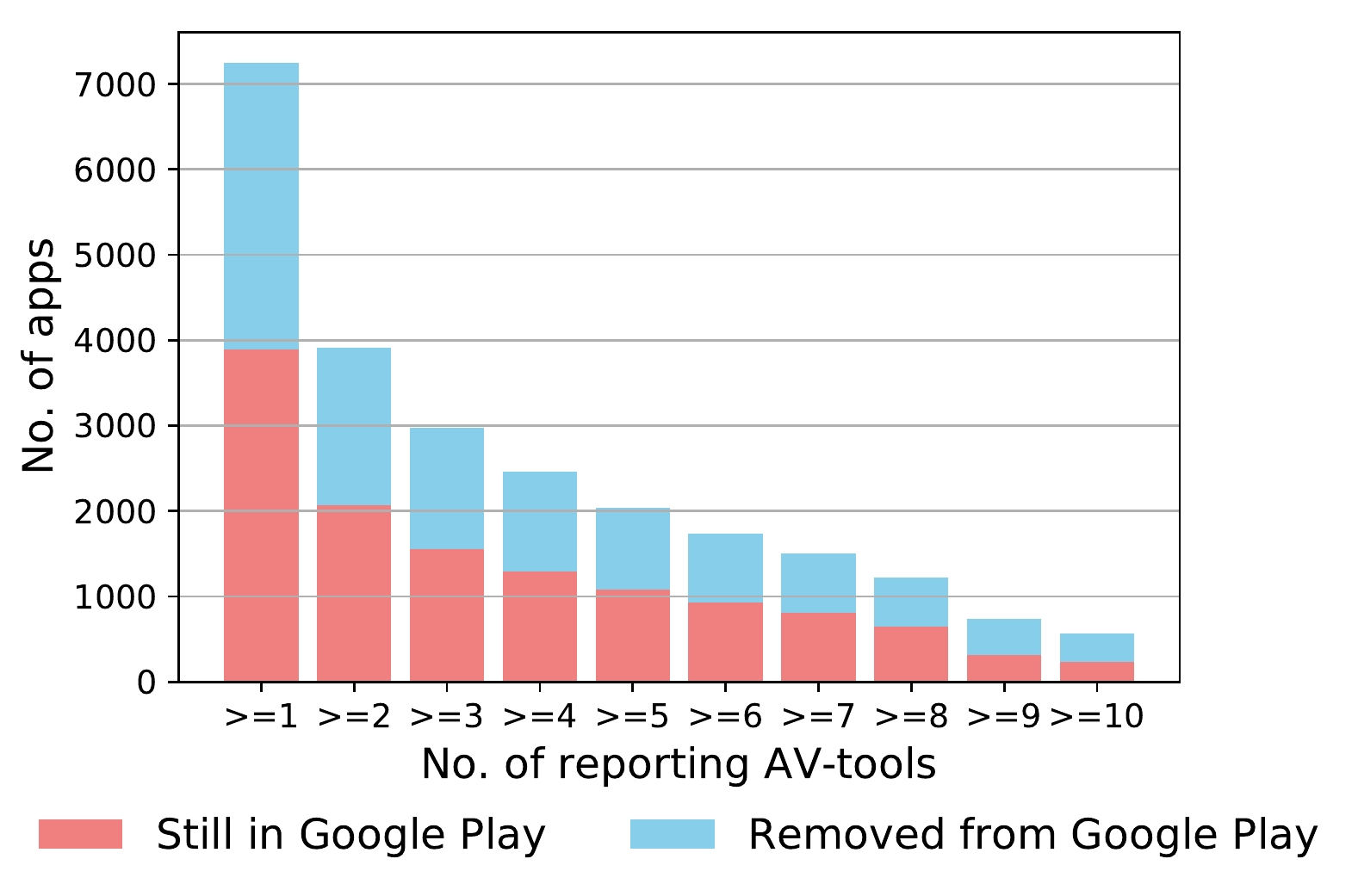}%
\caption{Number of apps against  the number of reporting AV-tools in VirusTotal}
\label{Fig:MalwareCount}
\end{figure}

In \figurename~\ref{Fig:MalwareCount}, we show a summarised view of the number of apps that were tagged for possible inclusion of malware by one or more AV-tools in VirusTotal and their availability in Google Play Store as of \emph{24-10-2018}. As the figure shows, there are 7,246 APKs that are tagged by at least one of the AV-tool.  

However, there can be false positives and as such a single AV-tool tagging an APK as malware in VirusTotal may not necessarily mean that the APK contains malware. As a result, previous work used different thresholds for the number of AV-tools that must report to consider an APK as malware. Ikram et al.~\cite{ikram2016analysis} used a conservative threshold of 5 and Arp et al.~\cite{arp2014drebin} used a more relaxed threshold of 2. \figurename~\ref{Fig:MalwareCount} shows that we have 3,907 apps if the  AV-tool threshold is 2 and 2,040 apps if the threshold is 5, out of which 2,067 and 1,080 apps respectively, are still there in Google Play Store. Approximately 46\% of the apps (3,358) that were tagged by at least one AV-tool are currently not available in Google Play Store. One possible reason is that Google at some point decided to remove those apps from the Play Store after receiving customer complaints or after post app publication binary analysis.

In Table~\ref{Tab:ExampleMalware}, we show some example apps that were tagged as containing malware, corresponding original apps, and their number of downloads. The table shows that while the counterfeit does not attract as large numbers of downloads as the original app in some occasions they have been downloaded significant number of times (e.g. Temple Theft Run has been downloaded at least 500,000 times.).

\begin{table}
\footnotesize{
\centering
\caption{Example similar apps that contain malware} \vspace{-3mm}
\begin{tabular}{p{1.4cm}|p{1.7cm}|p{0.6cm}|p{1.6cm}|p{1.5cm}}

\specialrule{.12em}{1em}{0em}
 \textbf{Original app} & \textbf{Similar \newline app} &   \center{\vspace*{-0.3cm}\textbf{AV tools}} & \textbf{Downloads (Original)} &  \textbf{Downloads (Similar)} \\ \hline
 \specialrule{.12em}{0em}{0em}
 
 & & & & \\
 \begin{minipage}{.3\textwidth}
\includegraphics[scale=0.07]{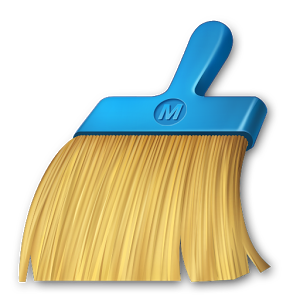} \\
\scriptsize{Clean  Master} \\  
\end{minipage} &  
        
\begin{minipage}{.3\textwidth}
\includegraphics[scale=0.07]{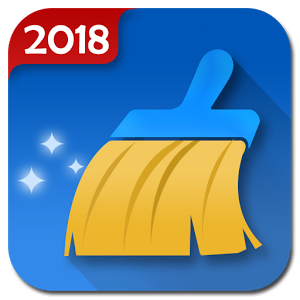} \\
\scriptsize{Ram Booster*} \\  
\end{minipage} &  
    
\center{12} & 500 million \newline - 1 billion & 500 \newline - 1,000 \\ 


 
 \begin{minipage}{.3\textwidth}
\includegraphics[scale=0.12]{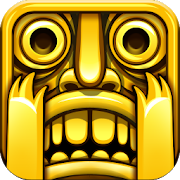} \\
\scriptsize{Temple Run} \\  
\end{minipage} &  
        
\begin{minipage}{.3\textwidth}
\includegraphics[scale=0.07]{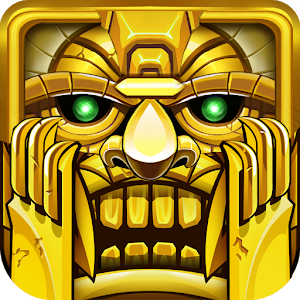} \\
\scriptsize{Endless Run*} \\  
\end{minipage} &  
    
\center{12} & 100 million \newline  - 500 million & 5,000 \newline - 10,000\\ 


 
 \begin{minipage}{.3\textwidth}
\includegraphics[scale=0.07]{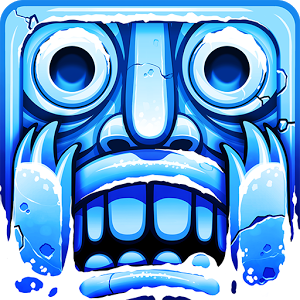} \\
\scriptsize{Temple Run 2} \\  
\end{minipage} &  
        
\begin{minipage}{.3\textwidth}
\includegraphics[scale=0.07]{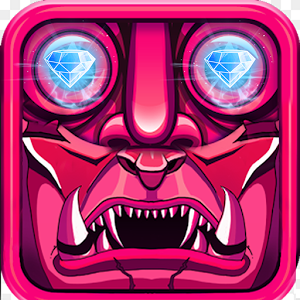} \\
\scriptsize{Temple Theft \newline Run*} \\ 
\end{minipage} &  
    
\center{12} & 500 million \newline  - 1  billion & 500,000 \newline - 1 million\\ 


 
 \begin{minipage}{.3\textwidth}
\includegraphics[scale=0.07]{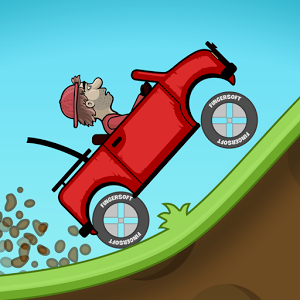} \\
\scriptsize{Hill Climb \newline Racing}  \\
\end{minipage} &  
        
\begin{minipage}{.3\textwidth}
\includegraphics[scale=0.07]{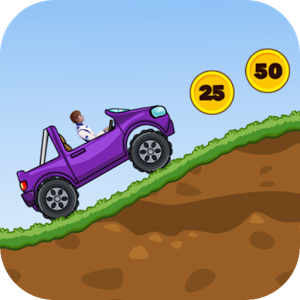} \\
\scriptsize{Offroad Racing: \newline Mountain Climb} \\  
\end{minipage} &  
    
\center{9} & 100 million \newline - 500 million & 1 million \newline - 5 million\\ 


 
 \begin{minipage}{.3\textwidth}
\includegraphics[scale=0.07]{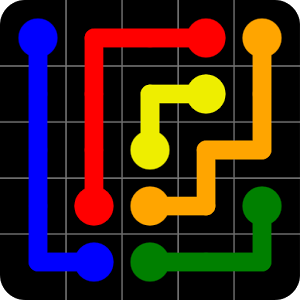} \\
\scriptsize{Flow Free} \\  
\end{minipage} &  
        
\begin{minipage}{.3\textwidth}
\includegraphics[scale=0.07]{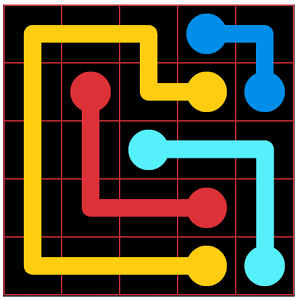} \\
\scriptsize{Colored Pipes} \\  
\end{minipage} &  
    
\center{8} & 100 million \newline - 500 million & 1 million \newline - 5 million\\ 


 
 \begin{minipage}{.3\textwidth}
\includegraphics[scale=0.07]{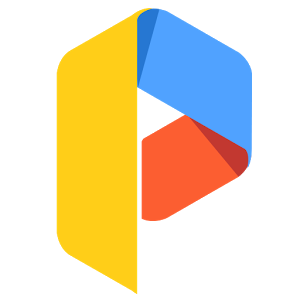} \\
\scriptsize{Parallel Space} \\  
\end{minipage} &  
        
\begin{minipage}{.3\textwidth}
\includegraphics[scale=0.08]{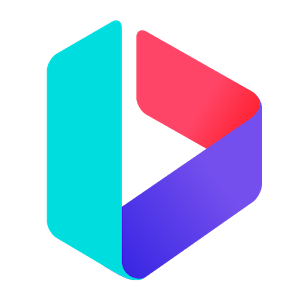} \\
\scriptsize{Double Account*} \\  
\end{minipage} &  
    
\center{17} & 50 million \newline - 100 million & 100,000 \newline - 500,000 \\ 



\specialrule{.12em}{0em}{0em}

\end{tabular}
\label{Tab:ExampleMalware}
\caption*{{\normalfont{ * The app is currently not available in Google Play Store}}}} \vspace{-3mm}
\end{table}

\subsection{Permission Requests}
\begin{table}
\footnotesize{
\centering
\caption{Example similar apps with high \emph{permission difference}} \vspace{-3mm}
\begin{tabular}{p{1.6cm}|p{1.6cm}|p{0.6cm}|p{1.6cm}|p{1.5cm}}
\specialrule{.12em}{1em}{0em}
 \textbf{Original app} & \textbf{Similar app} &   \textbf{\#PD} & \textbf{Downloads (Original)} &  \textbf{Downloads (Similar)} \\ \hline
 \specialrule{.12em}{0em}{0em}
 
 & & & & \\
 \begin{minipage}{.3\textwidth}
\includegraphics[scale=0.07]{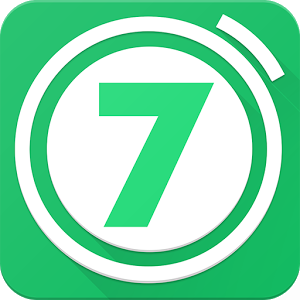} \\
\scriptsize{7 Minutes \newline Workout} \\  
\end{minipage} &  
        
\begin{minipage}{.3\textwidth}
\includegraphics[scale=0.075]{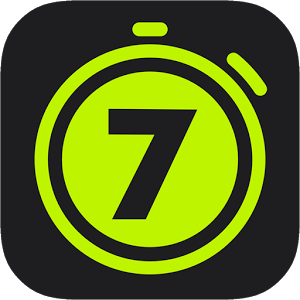} \\
\scriptsize{7 Minute \newline Workout VGFIT} \\  
\end{minipage} &  
    
\center{6} & 10 million \newline - 50 million & 5,000 \newline- 10,000 \\


 
 \begin{minipage}{.3\textwidth}
\includegraphics[scale=0.08]{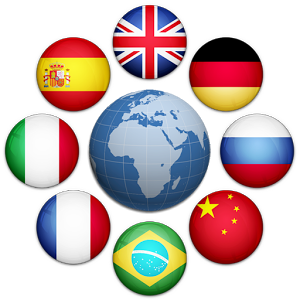} \\
\scriptsize{Language \newline Translator} \\  
\end{minipage} &  
        
\begin{minipage}{.3\textwidth}
\includegraphics[scale=0.08]{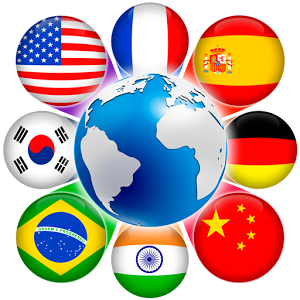} \\
\scriptsize{Multi Language \newline  Translator Free} \\  
\end{minipage} &  
    
\center{9} & 5 million \newline - 10 million & 100,000 \newline - 500,000\\ 


 
%
%


 
 \begin{minipage}{.3\textwidth}
\includegraphics[scale=0.07]{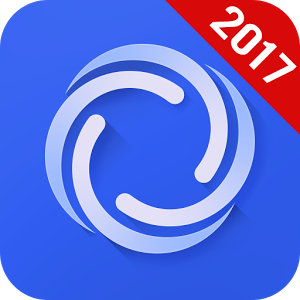} \\
\scriptsize{Phone Clean \newline  Speed Booster}  \\
\end{minipage} &  
        
\begin{minipage}{.3\textwidth}
\includegraphics[scale=0.08]{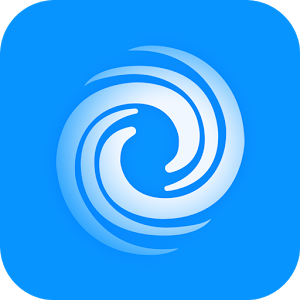} \\
\scriptsize{Lemon \newline Cleaner*} \\  
\end{minipage} &  
    
\center{12} & 1 million \newline- 5 million & 10,000 \newline - 50,000\\ 


 
 \begin{minipage}{.3\textwidth}
\includegraphics[scale=0.08]{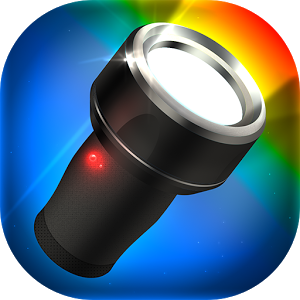} \\
\scriptsize{Color Torch HD \newline LED flash light} \\  
\end{minipage} &  
        
\begin{minipage}{.3\textwidth}
\includegraphics[scale=0.08]{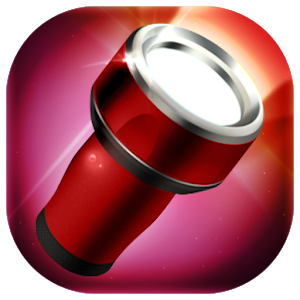} \\
\scriptsize{Flashlight \newline  Messenger*} \\  
\end{minipage} &  
    
\center{12} & 50 million \newline- 100 million & 1,000 \newline - 5,000\\ 


 
 \begin{minipage}{.3\textwidth}
\includegraphics[scale=0.08]{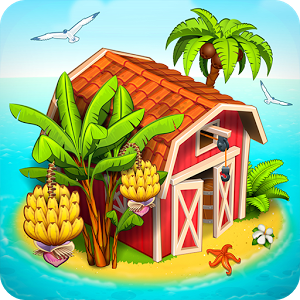} \\
\scriptsize{Farm Paradise: \newline Hay Island Bay} \\  
\end{minipage} &  
        
\begin{minipage}{.3\textwidth}
\includegraphics[scale=0.08]{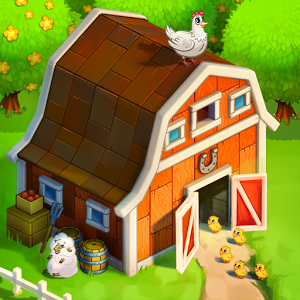} \\
\scriptsize{Summer \newline Tales} \\  
\end{minipage} &  
    
\center{23} & 1 million \newline - 5 million & 50,000 \newline - 100,000 \\ 


\begin{minipage}{.3\textwidth}
	\includegraphics[scale=0.08]{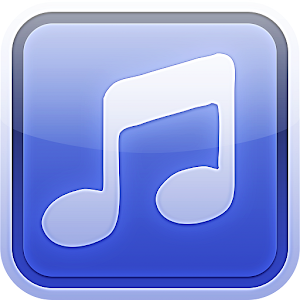} \\
	\scriptsize{Mp3 Music \newline Download} \\  
\end{minipage} &  

\begin{minipage}{.3\textwidth}
	\includegraphics[scale=0.08]{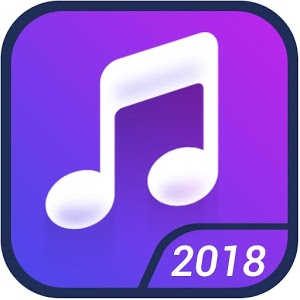} \\
	\scriptsize{Colorful Music \newline Player*} \\  
\end{minipage} &  

\center{5} & 5 million \newline - 10 million & 100,000 \newline - 500,000 \\ 


\specialrule{.12em}{0em}{0em}

\end{tabular}
\label{Tab:Permission}
\caption*{{\normalfont{ * The app is currently not available in Google Play Store \newline \#PD - Number of Permission differences}}}} \vspace{-3mm}
\end{table}

In addition to spreading malware, there can be other motivations for developing counterfeits. One such motivation can be collecting user's personal data by misleading them to install the counterfeit and requesting to grant \emph{dangerous Android permissions}. To investigate this, we considered the 26 dangerous permissions listed in the Android developer documentation~\cite{android}. To identify the potential counterfeits that ask for more permissions than the original app, we define a metric, {\bf\emph{permissions difference}}, which is the difference between the number of dangerous permissions requested by the potential counterfeit but not the original app and number of dangerous permissions requested by the original app but not by the potential counterfeit app. If the {\bf\emph{permissions difference}} is a positive value that means the potential counterfeit asks for more dangerous permissions than the original app and vice versa if the  {\bf\emph{permissions difference}} is negative. For the 49,608 potential counterfeits we had the APK files, we calculated the \emph{permission difference}. The permissions were extracted by the decompiling the APK and parsing the \emph{Android Manifest} file.

\begin{figure}
\centering 
\subfloat[Cumulative number of apps against \emph{permission difference}]{\label{Fig:CDFPermission}\includegraphics[trim=0cm 0cm 0cm 0cm, clip=true,scale=0.58]{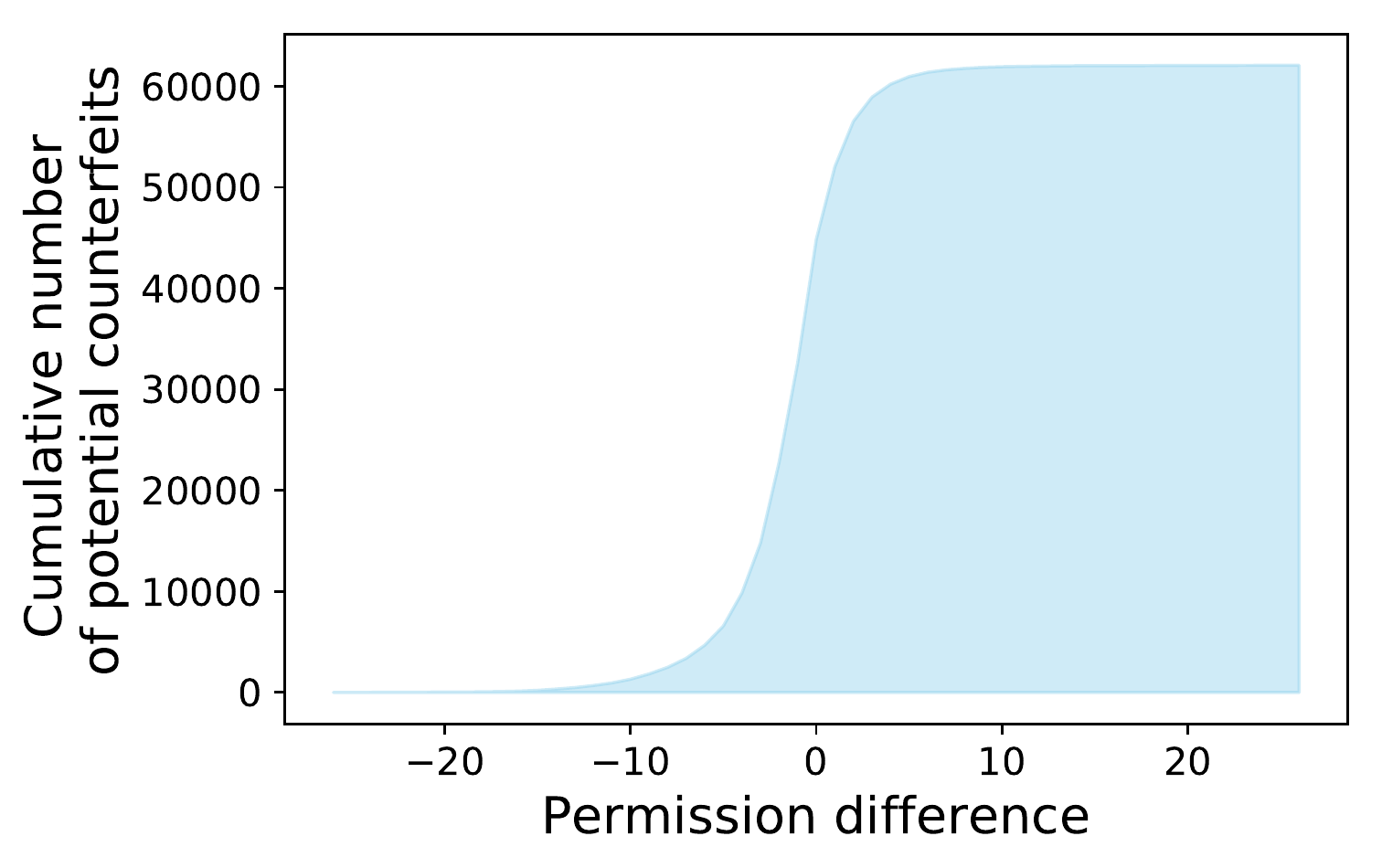}}\\
\subfloat[Play Store availability of apps with positive \emph{permission difference}]{\label{Fig:PSPermission}\includegraphics[trim=0cm 0cm 0cm 0cm, clip=true,scale=0.58]{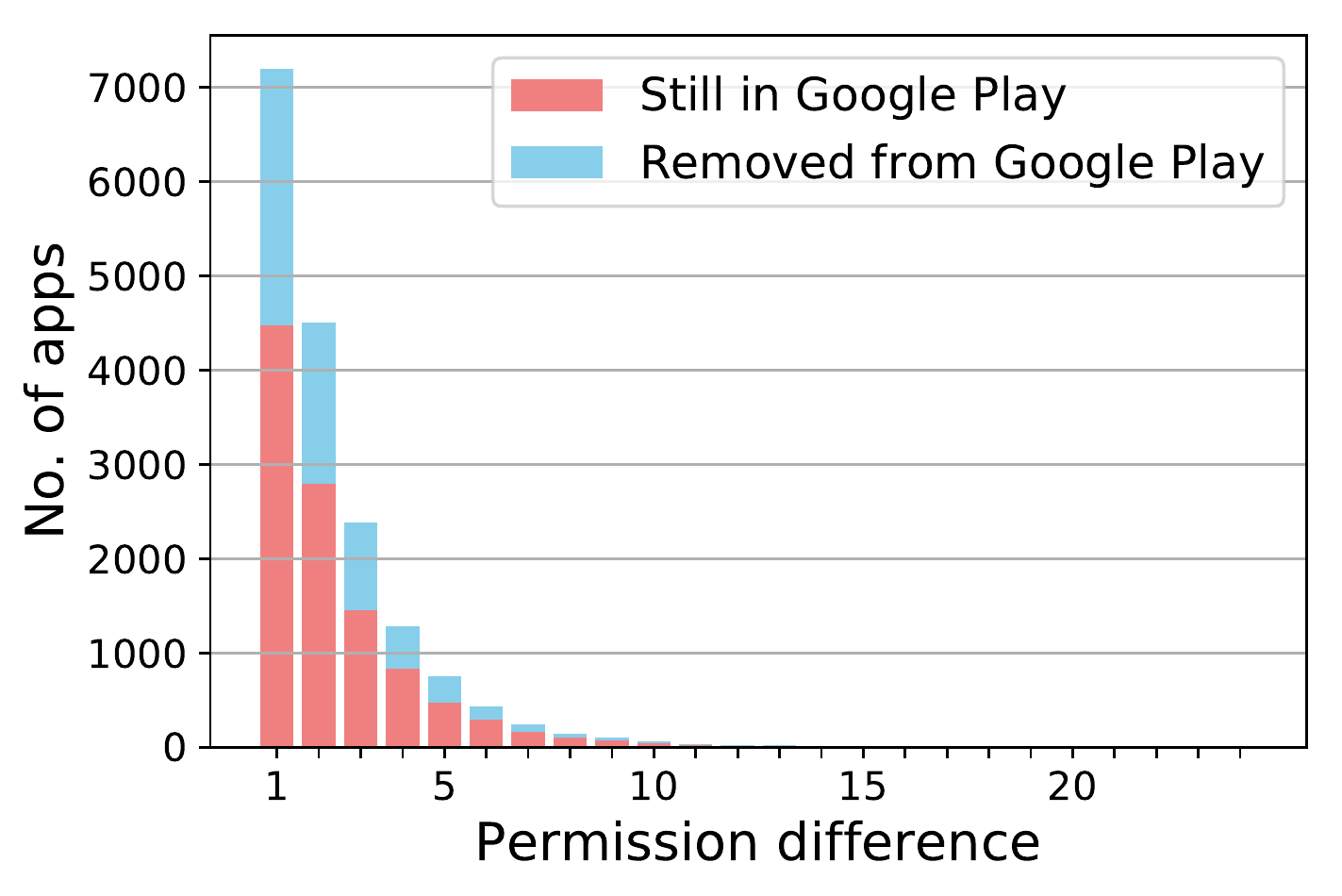}}
\caption{Potential counterfeits requesting additional dangerous permissions} \vspace{-6mm}
\label{Fig:Permissions}
\end{figure}

The cumulative sum of number of apps against the \emph{permission difference} is shown in Figure~\ref{Fig:CDFPermission}. The value of permission difference can vary from -26, where the counterfeit does not ask for any dangerous permission whereas the original app asks for all the dangerous permissions, to 26 for the opposite. Also, note that in this graph we have data for 62,074 apps instead of the 49,608 unique apps, because some apps were retrieved as counterfeits to more than one app in top-10,000 popular apps giving multiples values for permission difference. According to the figure, the majority of the potential counterfeits did not ask for more dangerous permissions than the original app. However, still there is 17,230 potential counterfeits that are asking at least one dangerous permission than the corresponding original app (13,857 unique apps), and 1,866 potential counterfeits (1,565 unique apps) asking at least five additional dangerous permissions compared to the original apps. 

In Table~\ref{Tab:Permission} we show some example such apps and in Figure~\ref{Fig:PSPermission} we show Google Play Store availability of the 17,230 apps that were asking for more dangerous permissions than the original app as of  \emph{24-10-2018}. As the figure shows approximately 37\% of the potential counterfeits with a permission difference of five is currently not available in the Google Play Store. Overall approximately 27\% of the apps with a positive permission difference are currently not available in Google Play Store. Again we conjecture these removals are done by Google after user complaints or post publication analysis. 

\subsection{Advertisement Libraries}
\begin{table}
\footnotesize{ 
\centering
\caption{Example similar apps with high \emph{ad library difference}} \vspace{-3mm}
\begin{tabular}{p{1.4cm}|p{1.7cm}|p{0.6cm}|p{1.6cm}|p{1.5cm}}
\specialrule{.12em}{1em}{0em}
 \textbf{Original app} & \textbf{Similar \newline app} &   \textbf{\#AD} & \textbf{Downloads (Original)} &  \textbf{Downloads (Similar)} \\ \hline
 \specialrule{.12em}{0em}{0em}
 
 & & & & \\
 
 \begin{minipage}{.3\textwidth}
\includegraphics[scale=0.12]{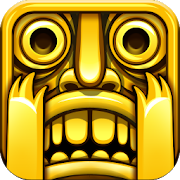} \\
\scriptsize{Temple Run} \\  
\end{minipage} &  
        
\begin{minipage}{.3\textwidth}
\includegraphics[scale=0.07]{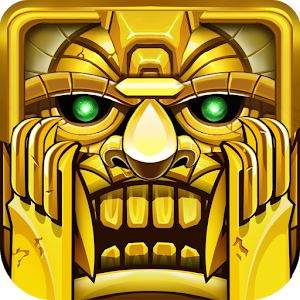} \\
\scriptsize{Endless Run*} \\  
\end{minipage} &  
    
\center{6} & 100 million \newline- 500 million & 5,000 \newline- 10,000 \\


 
 \begin{minipage}{.3\textwidth}
\includegraphics[scale=0.12]{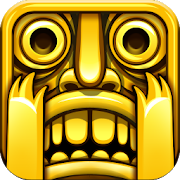} \\
\scriptsize{Temple Run} \\  
\end{minipage} &  
        
\begin{minipage}{.3\textwidth}
\includegraphics[scale=0.07]{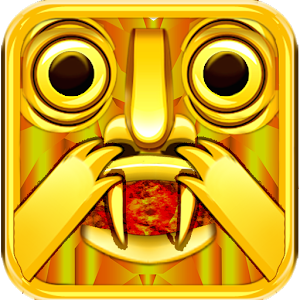} \\
\scriptsize{Temple escape*} \\  
\end{minipage} &  
    
\center{9} & 100 million \newline- 500 million & 50,000 \newline- 100,000\\ 


 
 \begin{minipage}{.3\textwidth}
\includegraphics[scale=0.07]{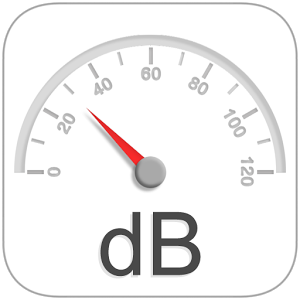} \\
\scriptsize{Sound \newline Meter} \\  
\end{minipage} &  
        
\begin{minipage}{.3\textwidth}
\includegraphics[scale=0.08]{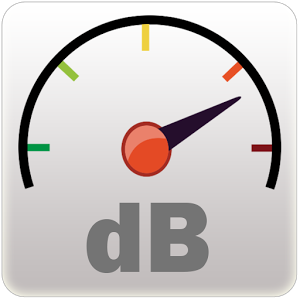} \\
\scriptsize{Smart Sound \newline Meter} \\ 
\end{minipage} &  
    
\center{6} & 5 million \newline- 10 million & 1,000 \newline- 5000\\ 


 
 \begin{minipage}{.3\textwidth}
\includegraphics[scale=0.07]{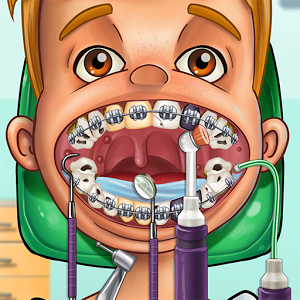} \\
\scriptsize{Dentist \newline games}  \\
\end{minipage} &  
        
\begin{minipage}{.3\textwidth}
\includegraphics[scale=0.075]{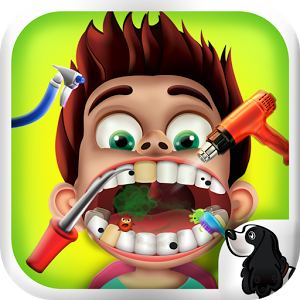} \\
\scriptsize{Dr. Dentist \newline Little*} \\  
\end{minipage} &  
    
\center{8} & 5 million \newline- 10 million & 100,000 \newline- 500,000\\ 


 
 \begin{minipage}{.3\textwidth}
\includegraphics[scale=0.07]{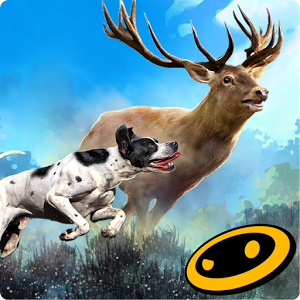} \\
\scriptsize{Deer Hunter \newline Classic} \\  
\end{minipage} &  
        
\begin{minipage}{.3\textwidth}
\includegraphics[scale=0.08]{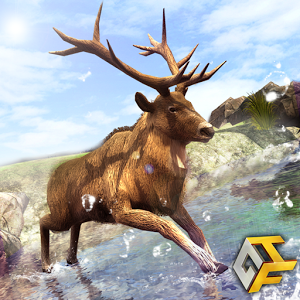} \\
\scriptsize{Sniper Deer \newline Hunting*} \\  
\end{minipage} &  
    
\center{8} & 50 million \newline- 100 million & 1,000 \newline-  5,000\\ 


 
%
%


\begin{minipage}{.3\textwidth}
	\includegraphics[scale=0.08]{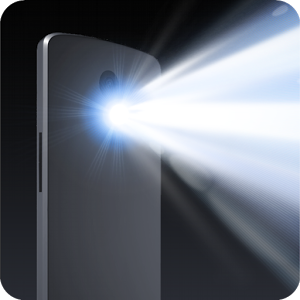} \\
	\scriptsize{Torch \newline Flashlight} \\  
\end{minipage} &  

\begin{minipage}{.3\textwidth}
	\includegraphics[scale=0.08]{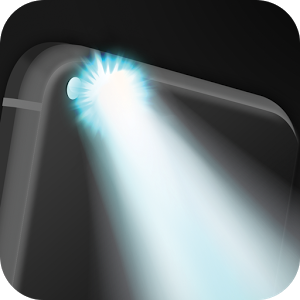} \\
	\scriptsize{Flashlight \newline Plus} \\  
\end{minipage} &  

\center{5} & 10 million  \newline-50 million & 10,000 \newline- 50,000 \\ 


\specialrule{.12em}{0em}{0em}

\end{tabular}
\label{Tab:ExampleAds}
\caption*{{\normalfont{ * The app is currently not available in Google Play Store \newline \#AD - Number of Ad library difference}}}} \vspace{-6mm}
\end{table}

\begin{figure}
\centering 
\subfloat[Cumulative number of apps against \emph{ad library difference}]{\label{Fig:CDFAds}\includegraphics[trim=0cm 0cm 0cm 0cm, clip=true,scale=0.58]{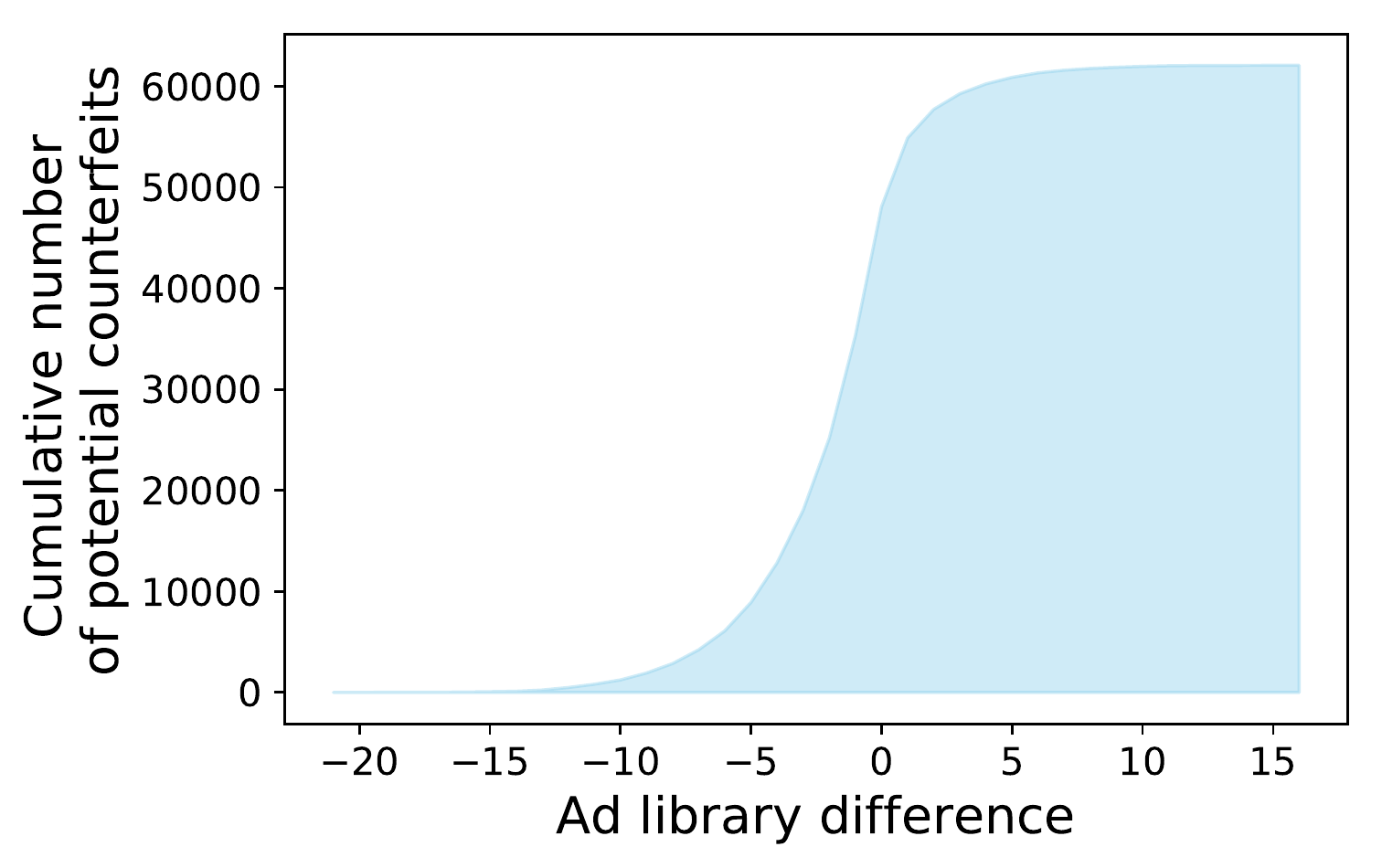}}\\
\subfloat[Play Store availability of apps with positive \emph{ad library difference}]{\label{Fig:PSAds}\includegraphics[trim=0cm 0cm 0cm 0cm, clip=true,scale=0.58]{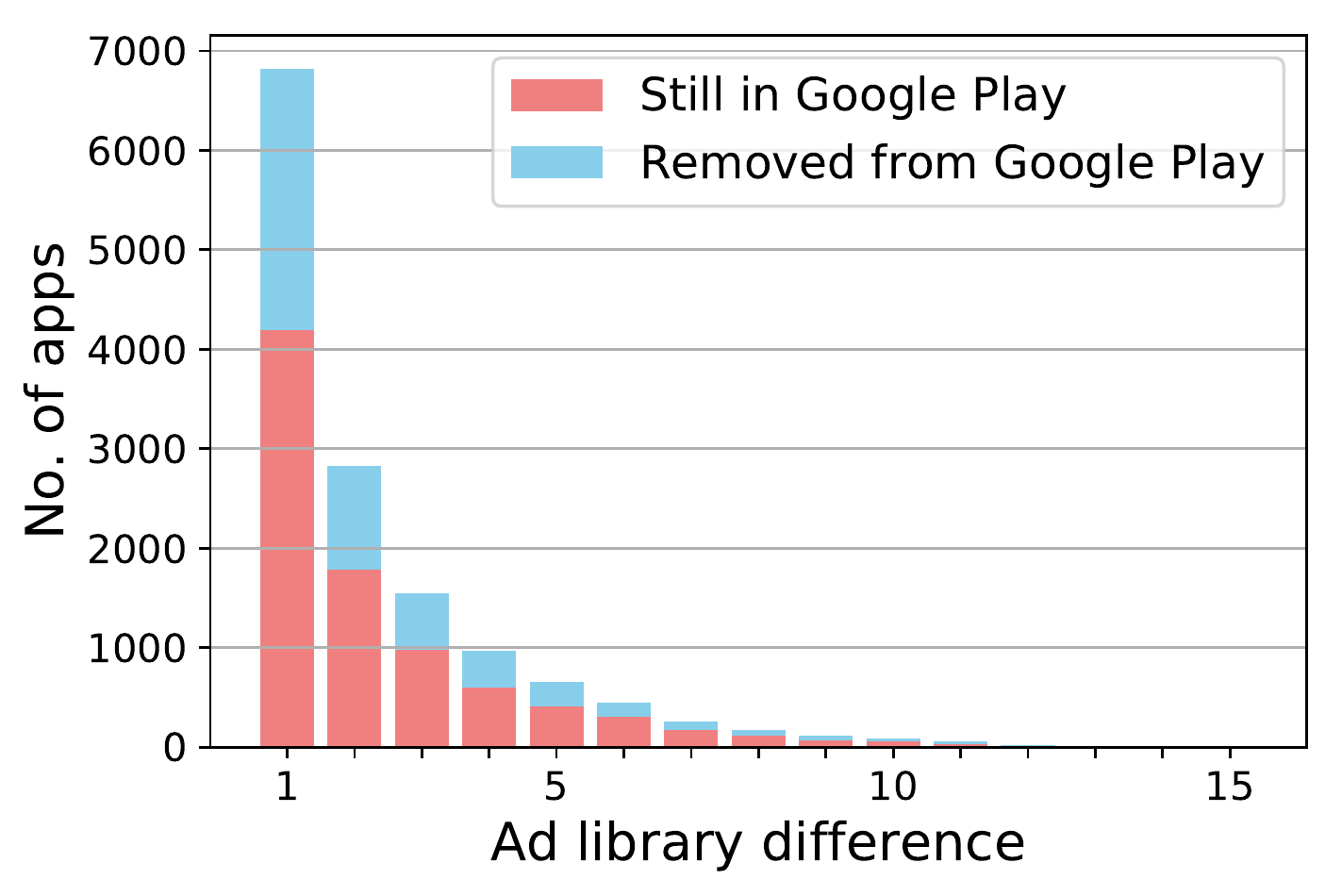}}
\caption{Potential counterfeits with additional third party advertisement libraries} \vspace{-4mm}
\label{Fig:Permissions}
\end{figure}

Another motivation behind developing counterfeits can be monetisation using third party advertisements and analytics. To quantify this, for each of the potential counterfeit we retrieved,  we defined a metric; {\bf\emph{ad library difference}} using the list of 124 mobile advertising and analytics libraries complied by Seneviratne et al.~\cite{seneviratne2015measurement}. Similar to previously calculated \emph{permission difference}, {\bf\emph{ad library difference}} is  the difference between the number of advertisement libraries embedded in the potential counterfeit but not in the original app and number of advertisement libraries embedded in the original app but not in the potential counterfeit app. A positive value of {\bf\emph{ad library difference}} means that the potential counterfeit has some extra ad libraries included compared to the original app. We show the cumulative number of potential counterfeits over the range of {\bf\emph{ad library difference}} in Figure ~\ref{Fig:CDFAds}. According to the figure, 13,997 apps (11,281 unique apps) have a positive ad library difference and out of that 1,841 (1,407 unique apps) have an ad library difference greater than or equal to five. Figure~\ref{Fig:PSAds} shows the Google Play store availability of apps with a positive ad library difference. Overall, approximately 33\% of the apps we identified are currently not available in the Google Play Store. \vspace{-3mm}

\section{Discussion}
\label{Sec:Discussion}

Using a large dataset of over 1.2 million app icons and over 1 million app executables, in this paper we presented insights of the app counterfeit problem in mobile app markets. The objective of the proposed embeddings-based method is to quickly and automatically assess a new submission and decide whether it resembles an existing app. If the new app is visually similar to an existing app, the app market operator can decide to do further security checks that can potentially include dynamic analysis as well as manual checks. We next discuss the limitations of our work and possible future extensions.

\subsection{Limitations}
Establishing a ground truth dataset for this type of problem is challenging due to several reasons. In this work, to build the ground truth dataset we used a heuristic approach to shortlist groups of apps that can potentially show visual similarities and then refine the dataset by manual inspection. However, as described in Section~\ref{Sec:Results}, there is a possibility that the unlabelled portion of data can still contain better similar apps than the labelled set and such apps can be returned during the nearest neighbour search instead of the target apps. This will result in a lower performance in terms of \emph{recall}; yet in reality received images also show high visual similarity. One possible solution for this is to use crowdsourcing to directly evaluate the performance of the embeddings without using a labelled dataset. For instance, retrieved images can be shown to a set of reviewers together with the original image and ask them to assign values for similarity with the original image. Then these values can be aggregated to come up with an overall score for the performance of the embedding. Crowdsourcing will also alleviate any biases introduced by individuals as visually similarity of images in some occasions can be subjective.

\textcolor{black}{\subsection{Adversarial Attacks}} 
\textcolor{black}{Many neural networks, especially CNN-based image classification systems are known to be prone to adversarial examples, i.e. curated inputs that can mislead the classifier to make a wrong decision~\cite{goodfellow2014explaining,kurakin2016adversarial}. In the context of image retrieval systems, an adversarial example is a query image that will retrieve irrelevant images. Since our work is based on an undefended pre-trained convolutional neural network, it is likely that an attacker can build adversarial examples that can bypass our similarity search and retrieve a set of unrelated images allowing the counterfeit apps to sustain in the app market. Multiple studies have demonstrated the feasibility of such attacks against image retrieval systems~\cite{szegedy2014intriguing,bai2020targeted,li2019universal}}.

\textcolor{black}{Also, we highlight that the vulnerability to adversarial attacks is not limited to deep learning based methods. Image retrieval systems based on traditional methods such as SIFT are also shown to be vulnerable to such attacks~\cite{hsu2009secure,do2010understanding}. Due to the multi-modal nature of our approach, an attacker might be able to curate images or text that can show high similarity in lower weighted modalities and bypass our system, yet appear as a real app to the user. It is an interesting future research direction to evaluate the robustness of our method to different adversarial attacks and integrate some defense mechanisms such as defensive distillation~\cite{papernot2016distillation} or projected gradient descent~\cite{madry2017towards} based methods. Also, to alleviate attacks that might rely on the relative balance of the weights of different modalities an ensemble of distance metrics can be considered with randomly perturbed weights.}

\subsection{Identifying Counterfeits}
\begin{figure}[h]
\centering
\includegraphics[scale=0.07]{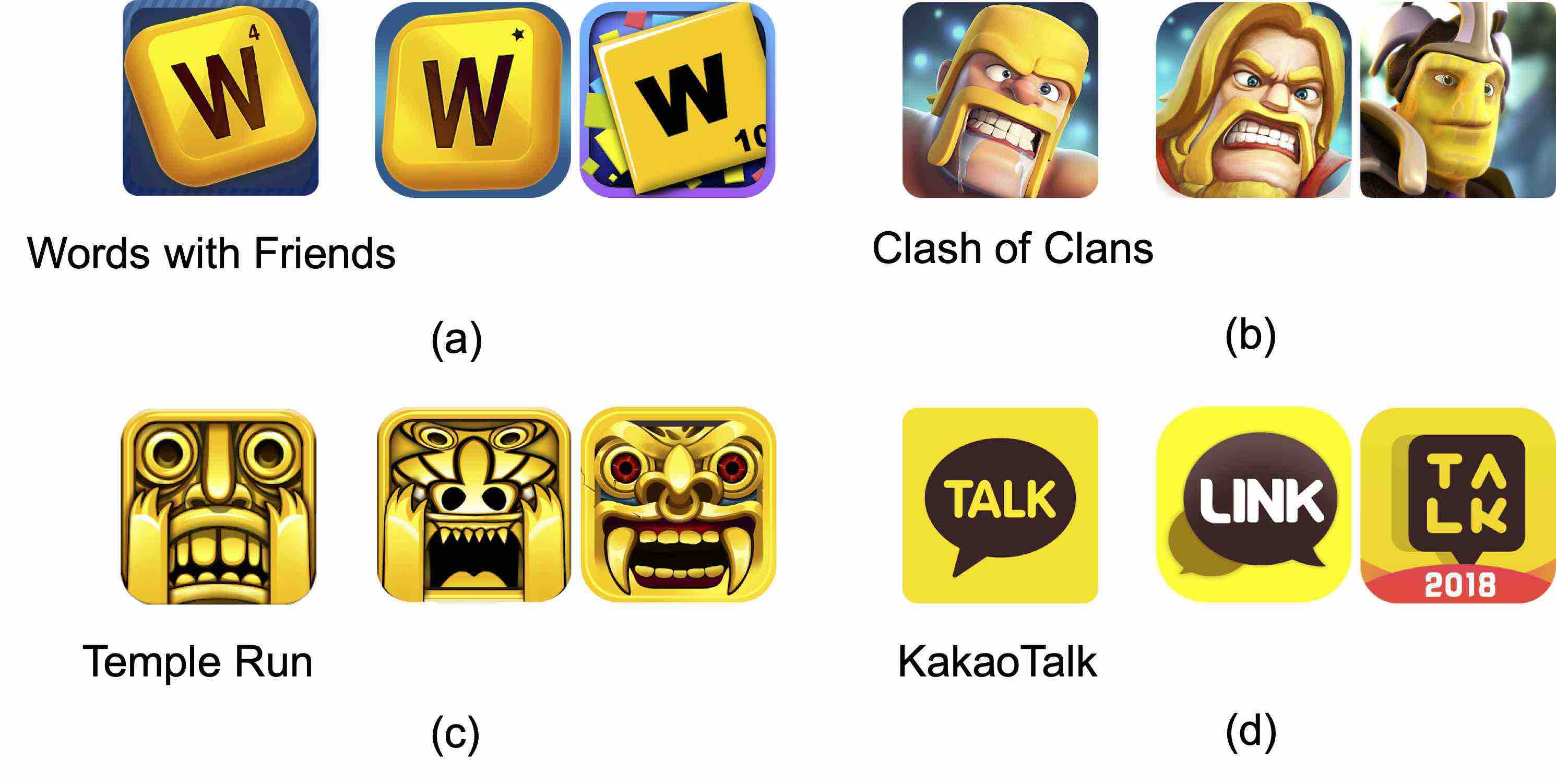}%
\caption{Some example apps that showed high visual similarity to one of the apps in top-10,000, yet did not contain any malware}
\label{Fig:NonMalware}
\end{figure}

Our evaluation of the retrieved set of highly visually similar apps was limited to the apps that possibly contain malware. Nonetheless, there can be counterfeits that do not contain malware and sometimes it can be difficult to  automatically decide whether a given app is counterfeit or not. In \figurename~\ref{Fig:NonMalware} we show some examples we retrieved that showed high visually similarity to one of the apps in top-1,000 yet did not contain malware or showed a significant difference in permissions or ad libraries. For instance, in \figurename~\ref{Fig:NonMalware}-(a) we show two visually similar apps we retrieved for  the popular game \emph{Words with Friends}. 

\textcolor{black}{One possible approach is to focus on apps that shows high visual and text similarity and high discrepancy in number of downloads. However, this is still not a straight forward decision. Firstly, there can be legitimate reasons for apps to be similar, at least in some modalities (e.g. generic apps such as messenger, flashlight, or phone). Moreover, developers can have multiple publishing profiles. If new apps were immediately counterfeited, the counterfeits can surpass the organic downloads using fake downloads~\cite{zhu2014discovery}. As a result, to detect counterfeits that does show any malware behaviours or privacy violations, some other factors such as app functionality and description need to be considered.}


For such scenarios, instead of the using only the similarity in app icons, the overall similarity of all the images available in Google Play Store pertaining to the two apps can be considered. This is because a developer can make the icons slightly different from the original app, and yet have the same visual \emph{``look and feel''} inside the app. Also, a number of work highlighted that apps can be clustered based on functional similarity using text mining methods~\cite{gorla2014checking,surian2017app}. Combining such methods with state-of-art techniques such as \emph{word vectors} and \emph{document vectors}~\cite{mikolov2013efficient,le2014distributed} and using them in conjunction with image similarity can further improve results. Nonetheless, for some cases still a manual intervention may be required. For example, in above case of \emph{Words with Friends} the similar apps are also word games and they are likely to show high visual similarity in internal GUIs as well as textual descriptions. In such scenarios again it might be useful to rely on crowdsourcing to obtain an overall assessment.

\section{Conclusion}
\label{Sec:Conclusion}

Using a dataset of over 1.2 million app icons, text descriptions, and their executables, we presented the problem of counterfeits in mobile app markets. We proposed an icon encoding method that allows to efficiently search potential counterfeits to a given app, using neural embeddings generated by a state-of-the-art convolutional neural network. More specifically, for app counterfeit detection problem, we showed that content and style neural embeddings generated from pre-trained VGGNet significantly outperforms hashing and feature-based image retrieval methods. We also showed that adding the text embeddings generated from the app descriptions further increase counterfeit detection rates.

To investigate the depth of the app counterfeit problem in Google Play Store, we used our multi-modal embedding methodology to retrieve potential counterfeits for the top-10,000 popular apps and investigated the possible inclusion of malware, permission usage, and embedded third party advertisement libraries. We found that 2,040 potential counterfeits we retrieved were marked by at least five commercial antivirus tools as malware, 1,565 asked for at least five additional dangerous permissions, and 1,407 had at least five additional embedded third party advertisement libraries. 

Finally, we showed that after 6-10 months since we discovered the apps, 27\%--46\% of the potential counterfeits we identified are not available in Google Play Store, potentially removed due to customer complaints or post publication findings. This is an indication that our proposed method is effective in identifying counterfeits at an early stage. We also showed that some of these apps are downloaded thousands of times before they are taken down. To sum up, our proposed multi-modal neural embedding approach allows to efficiently and effectively identify whether an app submitted by developer is trying counterfeit an existing popular app during the app publication process.

%

\section*{Acknowledgement}

This project is partially funded by the Google Faculty Rewards 2017, NSW Cyber Security Network's Pilot Grant Program 2018, and the Next Generation Technologies Program. Authors would like to thank VirusTotal for kindly providing access to the private API, which was used for the malware analysis in this paper.

\bibliographystyle{IEEEtran}
\bibliography{bibliography2}

\begin{thebibliography}{10}
\providecommand{\url}[1]{#1}
\csname url@samestyle\endcsname
\providecommand{\newblock}{\relax}
\providecommand{\bibinfo}[2]{#2}
\providecommand{\BIBentrySTDinterwordspacing}{\spaceskip=0pt\relax}
\providecommand{\BIBentryALTinterwordstretchfactor}{4}
\providecommand{\BIBentryALTinterwordspacing}{\spaceskip=\fontdimen2\font plus
\BIBentryALTinterwordstretchfactor\fontdimen3\font minus
  \fontdimen4\font\relax}
\providecommand{\BIBforeignlanguage}[2]{{%
\expandafter\ifx\csname l@#1\endcsname\relax
\typeout{** WARNING: IEEEtran.bst: No hyphenation pattern has been}%
\typeout{** loaded for the language `#1'. Using the pattern for}%
\typeout{** the default language instead.}%
\else
\language=\csname l@#1\endcsname
\fi
#2}}
\providecommand{\BIBdecl}{\relax}
\BIBdecl

\bibitem{rajasegaran2019multi}
J.~Rajasegaran, N.~Karunanayake, A.~Gunathillake, S.~Seneviratne, and
  G.~Jourjon, ``A multi-modal neural embeddings approach for detecting mobile
  counterfeit apps,'' in \emph{The World Wide Web Conference}.\hskip 1em plus
  0.5em minus 0.4em\relax ACM, 2019, pp. 3165--3171.

\bibitem{google2020}
``Number of available applications in the {G}oogle {P}lay store from {D}ecember
  2009 to {M}arch 2020,''
  https://www.statista.com/statistics/266210/number-of-available-applications-in-the-google-play-store/,
  {A}ccessed: 2020-05-116.

\bibitem{apple2020}
``How many apps are in the app store?''
  https://www.lifewire.com/how-many-apps-in-app-store-2000252, {A}ccessed:
  2020-05-116.

\bibitem{zhou2012dissecting}
Y.~Zhou and X.~Jiang, ``Dissecting android malware: {C}haracterization and
  evolution,'' in \emph{Security and Privacy (SP), 2012 IEEE Symposium
  on}.\hskip 1em plus 0.5em minus 0.4em\relax IEEE, 2012.

\bibitem{Seneviratne2015}
S.~Seneviratne, A.~Seneviratne, M.~A. Kaafar, A.~Mahanti, and P.~Mohapatra,
  ``Early detection of spam mobile apps,'' in \emph{Proceedings of the 24th
  International Conference on World Wide Web}, ser. WWW '15.\hskip 1em plus
  0.5em minus 0.4em\relax International World Wide Web Conferences Steering
  Committee, 2015.

\bibitem{whatsapp2017}
``Fake {W}hatsapp on {G}oogle {P}lay {S}tore downloaded by over 1 million
  {A}ndroid users,''
  https://thehackernews.com/2017/11/fake-whatsapp-android.html, {A}ccessed:
  2017-12-11.

\bibitem{angrybirds2012}
``Fake {A}ngry {B}irds {S}pace {A}ndroid app is full of malware,''
  https://www.gizmodo.com.au/2012/04/psa-fake-angry-birds-space-android-app-is-full-of-malware/,
  {A}ccessed: 2017-12-11.

\bibitem{sarah2013new}
S.~Perez, ``Developer spams {G}oogle {P}lay with ripoffs of well-known apps
  again,'' http://techcrunch.com, 2013.

\bibitem{netflix2017}
``Fake {N}etflix app takes control of {A}ndroid devices,''
  http://www.securityweek.com/fake-netflix-app-takes-control-android-devices,
  {A}ccessed: 2017-12-11.

\bibitem{cba2018}
``Scam alert: {F}ake {C}{B}{A} and {A}{N}{Z} bank apps discovered on {G}oogle
  {P}lay {S}tore,''
  https://www.lifehacker.com.au/2018/09/scam-alert-fake-cba-and-anz-banking-apps-found-on-google-play-store/,
  {A}ccessed: 2018-10-15.

\bibitem{securelist2018}
``Pocket cryptofarms: Investigating mobile apps for hidden mining,''
  https://securelist.com/pocket-cryptofarms/85137/.

\bibitem{crussell2013andarwin}
J.~Crussell, C.~Gibler, and H.~Chen, ``Andarwin: Scalable detection of
  semantically similar {A}ndroid applications,'' in \emph{European Symposium on
  Research in Computer Security}.\hskip 1em plus 0.5em minus 0.4em\relax
  Springer, 2013, pp. 182--199.

\bibitem{TempleRun}
``{T}emple {R}un,'' \url{https://tinyurl.com/celr2ff}, {A}ccessed: 2019-06-11.

\bibitem{alexNet}
A.~Krizhevsky, I.~Sutskever, and G.~E. Hinton, ``Imagenet classification with
  deep convolutional neural networks,'' in \emph{Advances in NIPS}, 2012.

\bibitem{vggNet}
K.~Simonyan and A.~Zisserman, ``Very deep convolutional networks for
  large-scale image recognition,'' \emph{CoRR}, vol. abs/1409.1556, 2014.

\bibitem{resNet}
K.~He, X.~Zhang, S.~Ren, and J.~Sun, ``Deep residual learning for image
  recognition,'' \emph{CoRR}, vol. abs/1512.03385, 2015.

\bibitem{SIFT}
D.~G. Lowe, ``Distinctive image features from scale-invariant keypoints,''
  \emph{International journal of computer vision}, vol.~60, no.~2, 2004.

\bibitem{SURF}
H.~Bay, T.~Tuytelaars, and L.~Van~Gool, ``{S}{U}{R}{F}: Speeded up robust
  features,'' in \emph{European conference on computer vision}.\hskip 1em plus
  0.5em minus 0.4em\relax Springer, 2006, pp. 404--417.

\bibitem{SSIM}
Z.~Wang, A.~C. Bovik, H.~R. Sheikh, and E.~P. Simoncelli, ``Image quality
  assessment: from error visibility to structural similarity,'' \emph{IEEE
  Transactions on Image Processing}, vol.~13, no.~4, pp. 600--612, 2004.

\bibitem{jegou2008hamming}
H.~Jegou, M.~Douze, and C.~Schmid, ``Hamming embedding and weak geometric
  consistency for large scale image search,'' in \emph{European conference on
  computer vision}.\hskip 1em plus 0.5em minus 0.4em\relax Springer, 2008, pp.
  304--317.

\bibitem{UKBench}
D.~Nister and H.~Stewenius, ``Scalable recognition with a vocabulary tree,'' in
  \emph{IEEE Computer Society Conference on Computer Vision and Pattern
  Recognition}, 2006.

\bibitem{le2014distributed}
Q.~Le and T.~Mikolov, ``Distributed representations of sentences and
  documents,'' in \emph{International conference on machine learning}, 2014,
  pp. 1188--1196.

\bibitem{grace2012riskranker}
M.~Grace, Y.~Zhou, Q.~Zhang, S.~Zou, and X.~Jiang, ``Riskranker: {S}calable and
  accurate zero-day {A}ndroid malware detection,'' in \emph{Proc. of the 10th
  international conference on Mobile systems, applications, and
  services}.\hskip 1em plus 0.5em minus 0.4em\relax ACM, 2012, pp. 281--294.

\bibitem{wu2012droidmat}
D.-J. Wu, C.-H. Mao, T.-E. Wei, H.-M. Lee, and K.-P. Wu, ``Droidmat: {A}ndroid
  malware detection through manifest and api calls tracing,'' in
  \emph{Information Security (Asia JCIS), 2012 Seventh Asia Joint Conference
  on}.\hskip 1em plus 0.5em minus 0.4em\relax IEEE, 2012, pp. 62--69.

\bibitem{burguera2011crowdroid}
I.~Burguera, U.~Zurutuza, and S.~Nadjm-Tehrani, ``Crowdroid: {B}ehavior-based
  malware detection system for android,'' in \emph{Proc. of the 1st ACM
  workshop on Security and privacy in smartphones and mobile devices}.\hskip
  1em plus 0.5em minus 0.4em\relax ACM, 2011, pp. 15--26.

\bibitem{shabtai2012andromaly}
A.~Shabtai, U.~Kanonov, Y.~Elovici, C.~Glezer, and Y.~Weiss, ``{A}ndromaly: {A}
  behavioral malware detection framework for android devices,'' \emph{Journal
  of Intelligent Information Systems}, vol.~38, 2012.

\bibitem{yuan2014droid}
Z.~Yuan, Y.~Lu, Z.~Wang, and Y.~Xue, ``Droid-{S}ec: {D}eep learning in
  {A}ndroid malware detection,'' in \emph{ACM SIGCOMM Computer Communication
  Review}, 2014.

\bibitem{xie2015appwatcher}
Z.~Xie and S.~Zhu, ``Appwatcher: Unveiling the underground market of trading
  mobile app reviews,'' in \emph{Proc. of the 8th ACM Conference on Security \&
  Privacy in Wireless and Mobile Networks}.\hskip 1em plus 0.5em minus
  0.4em\relax ACM, 2015.

\bibitem{chandy2012identifying}
R.~Chandy and H.~Gu, ``Identifying spam in the i{O}{S} app store,'' in
  \emph{Proc. of the 2nd Joint WICOW/AIRWeb Workshop on Web Quality}, 2012.

\bibitem{gibler2013adrob}
C.~Gibler, R.~Stevens, J.~Crussell, H.~Chen, H.~Zang, and H.~Choi, ``Adrob:
  Examining the landscape and impact of android application plagiarism,'' in
  \emph{Proceeding of the 11th annual international conference on Mobile
  systems, applications, and services}, 2013, pp. 431--444.

\bibitem{surian2017app}
D.~Surian, S.~Seneviratne, A.~Seneviratne, and S.~Chawla, ``App
  miscategorization detection: A case study on google play,'' \emph{IEEE
  T{K}{D}{E}}, vol.~29, no.~8, 2017.

\bibitem{seneviratne2017spam}
S.~Seneviratne, A.~Seneviratne, M.~A. Kaafar, A.~Mahanti, and P.~Mohapatra,
  ``Spam mobile apps: Characteristics, detection, and in the wild analysis,''
  \emph{ACM Transactions on the Web (TWEB)}, vol.~11, no.~1, pp. 1--29, 2017.

\bibitem{viennot2014measurement}
N.~Viennot, E.~Garcia, and J.~Nieh, ``A measurement study of {G}oogle {P}lay,''
  in \emph{ACM SIGMETRICS Performance Evaluation Review}, 2014.

\bibitem{sun2015droideagle}
M.~Sun, M.~Li, and J.~C. Lui, ``Droideagle: seamless detection of visually
  similar android apps,'' in \emph{Proceedings of the 8th ACM Conference on
  Security \& Privacy in Wireless and Mobile Networks}, 2015, pp. 1--12.

\bibitem{malisa2016mobile}
L.~Malisa, K.~Kostiainen, M.~Och, and S.~Capkun, ``Mobile application
  impersonation detection using dynamic user interface extraction,'' in
  \emph{European Symposium on Research in Computer Security}, 2016.

\bibitem{andow2016study}
B.~Andow, A.~Nadkarni, B.~Bassett, W.~Enck, and T.~Xie, ``A study of grayware
  on {G}oogle {P}lay,'' in \emph{Security and Privacy Workshops (SPW), 2016
  IEEE}.\hskip 1em plus 0.5em minus 0.4em\relax IEEE, 2016.

\bibitem{Malisa:2017}
L.~Malisa, K.~Kostiainen, and S.~Capkun, ``Detecting mobile application
  spoofing attacks by leveraging user visual similarity perception,'' in
  \emph{Proc. of the Seventh ACM on Conference on Data and Application Security
  and Privacy}, ser. CODASPY '17.\hskip 1em plus 0.5em minus 0.4em\relax New
  York, NY, USA: ACM, 2017.

\bibitem{gatys2015neural}
L.~A. Gatys, A.~S. Ecker, and M.~Bethge, ``A neural algorithm of artistic
  style,'' \emph{arXiv preprint arXiv:1508.06576}, 2015.

\bibitem{gatys2016image}
------, ``Image style transfer using convolutional neural networks,'' in
  \emph{Proceedings of the IEEE conference on computer vision and pattern
  recognition}, 2016, pp. 2414--2423.

\bibitem{jing2017neural}
Y.~Jing, Y.~Yang, Z.~Feng, J.~Ye, and M.~Song, ``Neural style transfer: {A}
  review,'' \emph{arXiv preprint arXiv:1705.04058}, 2017.

\bibitem{johnson2016perceptual}
J.~Johnson, A.~Alahi, and L.~Fei-Fei, ``Perceptual losses for real-time style
  transfer and super-resolution,'' in \emph{European conference on computer
  vision}.\hskip 1em plus 0.5em minus 0.4em\relax Springer, 2016, pp. 694--711.

\bibitem{bell2015siamese}
S.~Bell and K.~Bala, ``Learning visual similarity for product design with
  convolutional neural networks,'' \emph{ACM Transactions on Graphics (TOG)},
  2015.

\bibitem{tan2016ceci}
W.~R. Tan, C.~S. Chan, H.~E. Aguirre, and K.~Tanaka, ``Ceci n'est pas une pipe:
  A deep convolutional network for fine-art paintings classification,'' in
  \emph{Image Processing (ICIP), 2016 IEEE International Conference on}.

\bibitem{style_classification}
S.~Matsuo and K.~Yanai, ``{C}{N}{N}-based style vector for style image
  retrieval,'' in \emph{Proceedings of the 2016 ACM on International Conference
  on Multimedia Retrieval}, 2016, pp. 309--312.

\bibitem{seneviratne2015early}
S.~Seneviratne, A.~Seneviratne, M.~A. Kaafar, A.~Mahanti, and P.~Mohapatra,
  ``Early detection of spam mobile apps,'' in \emph{Proc. of the 24th
  International Conference on World Wide Web}, 2015.

\bibitem{sharif2014cnn}
A.~Sharif~Razavian, H.~Azizpour, J.~Sullivan, and S.~Carlsson, ``Cnn features
  off-the-shelf: an astounding baseline for recognition,'' in \emph{Proceedings
  of the IEEE conference on computer vision and pattern recognition workshops},
  2014, pp. 806--813.

\bibitem{jegou2010aggregating}
H.~J{\'e}gou, M.~Douze, C.~Schmid, and P.~P{\'e}rez, ``Aggregating local
  descriptors into a compact image representation,'' in \emph{2010 IEEE
  computer society conference on computer vision and pattern
  recognition}.\hskip 1em plus 0.5em minus 0.4em\relax IEEE, 2010, pp.
  3304--3311.

\bibitem{gong2012iterative}
Y.~Gong, S.~Lazebnik, A.~Gordo, and F.~Perronnin, ``Iterative quantization: A
  procrustean approach to learning binary codes for large-scale image
  retrieval,'' \emph{IEEE transactions on pattern analysis and machine
  intelligence}, vol.~35, no.~12, pp. 2916--2929, 2012.

\bibitem{philbin2007object}
J.~Philbin, O.~Chum, M.~Isard, J.~Sivic, and A.~Zisserman, ``Object retrieval
  with large vocabularies and fast spatial matching,'' in \emph{2007 IEEE
  conference on computer vision and pattern recognition}.\hskip 1em plus 0.5em
  minus 0.4em\relax IEEE, 2007, pp. 1--8.

\bibitem{philbin2008lost}
------, ``Lost in quantization: Improving particular object retrieval in large
  scale image databases,'' in \emph{2008 IEEE conference on computer vision and
  pattern recognition}.\hskip 1em plus 0.5em minus 0.4em\relax IEEE, 2008, pp.
  1--8.

\bibitem{arandjelovic2011smooth}
R.~Arandjelovi{\'c} and A.~Zisserman, ``Smooth object retrieval using a bag of
  boundaries,'' in \emph{2011 International Conference on Computer
  Vision}.\hskip 1em plus 0.5em minus 0.4em\relax IEEE, 2011, pp. 375--382.

\bibitem{L2_distance_not_good_for_high_diamentions}
C.~C. Aggarwal, A.~Hinneburg, and D.~A. Keim, ``On the surprising behavior of
  distance metrics in high dimensional spaces,'' in \emph{ICDT}.\hskip 1em plus
  0.5em minus 0.4em\relax Springer, 2001.

\bibitem{imagenet_cvpr09}
J.~Deng, W.~Dong, R.~Socher, L.-J. Li, K.~Li, and L.~Fei-Fei, ``Imagenet: A
  large-scale hierarchical image database,'' in \emph{2009 IEEE conference on
  computer vision and pattern recognition}.\hskip 1em plus 0.5em minus
  0.4em\relax IEEE, 2009, pp. 248--255.

\bibitem{fc7_is_good}
A.~Babenko, A.~Slesarev, A.~Chigorin, and V.~S. Lempitsky, ``Neural codes for
  image retrieval,'' \emph{CoRR}, vol. abs/1404.1777, 2014.

\bibitem{gram}
L.~A. Gatys, A.~S. Ecker, and M.~Bethge, ``Texture synthesis and the controlled
  generation of natural stimuli using convolutional neural networks,''
  \emph{CoRR}, vol. abs/1505.07376, 2015.

\bibitem{very_sparse_random_projection}
P.~Li, T.~J. Hastie, and K.~W. Church, ``Very sparse random projections,'' in
  \emph{Proceedings of the 12th ACM SIGKDD international conference on
  Knowledge discovery and data mining}, 2006, pp. 287--296.

\bibitem{bingham2001random}
E.~Bingham and H.~Mannila, ``Random projection in dimensionality reduction:
  applications to image and text data,'' in \emph{Proceedings of the seventh
  ACM SIGKDD international conference on Knowledge discovery and data mining},
  2001, pp. 245--250.

\bibitem{sparse_random_projection_orthogonal}
D.~Achlioptas, ``Database-friendly random projections,'' in \emph{Proceedings
  of the twentieth ACM SIGMOD-SIGACT-SIGART symposium on Principles of database
  systems}, 2001, pp. 274--281.

\bibitem{sparse_random_projection_orthogonal2}
V.~Sulic, J.~Per{\v{s}}, M.~Kristan, and S.~Kovacic, ``Efficient dimensionality
  reduction using random projection,'' in \emph{15th Computer Vision Winter
  Workshop}, 2010.

\bibitem{novak2016improving}
R.~Novak and Y.~Nikulin, ``Improving the neural algorithm of artistic style,''
  \emph{arXiv preprint arXiv:1605.04603}, 2016.

\bibitem{li2017demystifying}
Y.~Li, N.~Wang, J.~Liu, and X.~Hou, ``Demystifying neural style transfer,''
  \emph{arXiv preprint arXiv:1701.01036}, 2017.

\bibitem{rajasegaran2018neural}
J.~Rajasegaran, S.~Seneviratne, and G.~Jourjon, ``A neural embeddings approach
  for detecting mobile counterfeit apps,'' \emph{arXiv preprint
  arXiv:1804.09882}, 2018.

\bibitem{AHash}
``{L}ook {L}ike {I}t,''
  http://www.hackerfactor.com/blog/?/archives/432-Looks-Like-It.html,
  {A}ccessed: 2017-08-19.

\bibitem{DHash}
``{K}ind of {L}ike {T}hat,''
  http://www.hackerfactor.com/blog/?/archives/529-Kind-of-Like-That.html,
  {A}ccessed: 2018-08-19.

\bibitem{PHash}
H.~Zhang, M.~Schmucker, and X.~Niu, ``The design and application of phabs: A
  novel benchmark platform for perceptual hashing algorithms,'' in \emph{IEEE
  International Conference on Multimedia and Expo}, 2007.

\bibitem{WHash}
F.~Ahmed and M.~Y. Siyal, ``A secure and robust wavelet-based hashing scheme
  for image authentication,'' in \emph{Advances in Multimedia Modeling}, 2006,
  pp. 51--62.

\bibitem{AKAZE}
P.~F. Alcantarilla, J.~Nuevo, and A.~Bartoli, ``Fast explicit diffusion for
  accelerated features in nonlinear scale spaces,'' in \emph{BMVC}, 2013, pp.
  1--9.

\bibitem{LATCH}
G.~Levi and T.~Hassner, ``{LATCH}: Learned arrangements of three patch codes,''
  in \emph{Winter Conference on Applications of Computer Vision}, 2016.

\bibitem{satopaa2011finding}
V.~Satopaa, J.~Albrecht, D.~Irwin, and B.~Raghavan, ``Finding a" kneedle" in a
  haystack: Detecting knee points in system behavior,'' in \emph{Distributed
  Computing Systems Workshops (ICDCSW), 2011 31st International Conference
  on}.\hskip 1em plus 0.5em minus 0.4em\relax IEEE, 2011.

\bibitem{ikram2016analysis}
M.~Ikram, N.~Vallina-Rodriguez, S.~Seneviratne, M.~A. Kaafar, and V.~Paxson,
  ``An analysis of the privacy and security risks of android vpn
  permission-enabled apps,'' in \emph{Proceedings of the 2016 Internet
  Measurement Conference}, 2016, pp. 349--364.

\bibitem{arp2014drebin}
D.~Arp, M.~Spreitzenbarth, M.~Hubner, H.~Gascon, K.~Rieck, and C.~Siemens,
  ``Drebin: Effective and explainable detection of android malware in your
  pocket.'' in \emph{NDSS}, vol.~14, 2014, pp. 23--26.

\bibitem{android}
``Permissions overview,'' \url{https://tinyurl.com/y3ehq9mu}, {A}ccessed:
  2018-10-31.

\bibitem{seneviratne2015measurement}
S.~Seneviratne, H.~Kolamunna, and A.~Seneviratne, ``A measurement study of
  tracking in paid mobile applications,'' in \emph{Proceedings of the 8th ACM
  Conference on Security \& Privacy in Wireless and Mobile Networks}, 2015, pp.
  1--6.

\bibitem{goodfellow2014explaining}
I.~J. Goodfellow, J.~Shlens, and C.~Szegedy, ``Explaining and harnessing
  adversarial examples,'' \emph{arXiv preprint arXiv:1412.6572}, 2014.

\bibitem{kurakin2016adversarial}
A.~Kurakin, I.~Goodfellow, and S.~Bengio, ``Adversarial examples in the
  physical world,'' \emph{arXiv preprint arXiv:1607.02533}, 2016.

\bibitem{szegedy2014intriguing}
C.~Szegedy, W.~Zaremba, I.~Sutskever, J.~B. Estrach, D.~Erhan, I.~Goodfellow,
  and R.~Fergus, ``Intriguing properties of neural networks,'' in \emph{2nd
  International Conference on Learning Representations, ICLR 2014}, 2014.

\bibitem{bai2020targeted}
J.~Bai, B.~Chen, Y.~Li, D.~Wu, W.~Guo, S.-t. Xia, and E.-h. Yang, ``Targeted
  attack for deep hashing based retrieval,'' \emph{arXiv preprint
  arXiv:2004.07955}, 2020.

\bibitem{li2019universal}
J.~Li, R.~Ji, H.~Liu, X.~Hong, Y.~Gao, and Q.~Tian, ``Universal perturbation
  attack against image retrieval,'' in \emph{Proceedings of the IEEE
  International Conference on Computer Vision}, 2019, pp. 4899--4908.

\bibitem{hsu2009secure}
C.-Y. Hsu, C.-S. Lu, and S.-C. Pei, ``Secure and robust {S}{I}{F}{T},'' in
  \emph{Proceedings of the 17th ACM international conference on Multimedia},
  2009, pp. 637--640.

\bibitem{do2010understanding}
T.-T. Do, E.~Kijak, T.~Furon, and L.~Amsaleg, ``Understanding the security and
  robustness of {S}{I}{F}{T},'' in \emph{Proceedings of the 18th ACM
  international conference on Multimedia}, 2010, pp. 1195--1198.

\bibitem{papernot2016distillation}
N.~Papernot, P.~McDaniel, X.~Wu, S.~Jha, and A.~Swami, ``Distillation as a
  defense to adversarial perturbations against deep neural networks,'' in
  \emph{2016 IEEE Symposium on Security and Privacy (SP)}.\hskip 1em plus 0.5em
  minus 0.4em\relax IEEE, 2016, pp. 582--597.

\bibitem{madry2017towards}
A.~Madry, A.~Makelov, L.~Schmidt, D.~Tsipras, and A.~Vladu, ``Towards deep
  learning models resistant to adversarial attacks,'' \emph{arXiv preprint
  arXiv:1706.06083}, 2017.

\bibitem{zhu2014discovery}
H.~Zhu, H.~Xiong, Y.~Ge, and E.~Chen, ``Discovery of ranking fraud for mobile
  apps,'' \emph{IEEE Transactions on knowledge and data engineering}, vol.~27,
  no.~1, pp. 74--87, 2014.

\bibitem{gorla2014checking}
A.~Gorla, I.~Tavecchia, F.~Gross, and A.~Zeller, ``Checking app behavior
  against app descriptions,'' in \emph{Proceedings of the 36th International
  Conference on Software Engineering}, 2014, pp. 1025--1035.

\bibitem{mikolov2013efficient}
T.~Mikolov, K.~Chen, G.~Corrado, and J.~Dean, ``Efficient estimation of word
  representations in vector space,'' \emph{arXiv preprint arXiv:1301.3781},
  2013.

\end{thebibliography}
\vspace{-15mm}
\begin{IEEEbiography}[{\includegraphics[width=1in,height=1.25in,clip,keepaspectratio]{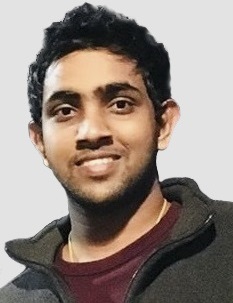}}]{Naveen Karunanayake}
	received his Bachelors degree in Electronic and Telecommunication Engineering (First Class Hons.) from University of Moratuwa, Sri Lanka in 2020. He is currently working as a Graduate Research Scholar at University of Moratuwa. He worked as a Visiting Research Intern at Data61-CSIRO, Sydney in 2018. His research interests include Machine learning, Computer vision and Algorithm development and optimization.
\end{IEEEbiography}

\vskip -3\baselineskip plus -1fil

\begin{IEEEbiography}[{\includegraphics[width=1in,height=1.25in,clip,keepaspectratio]{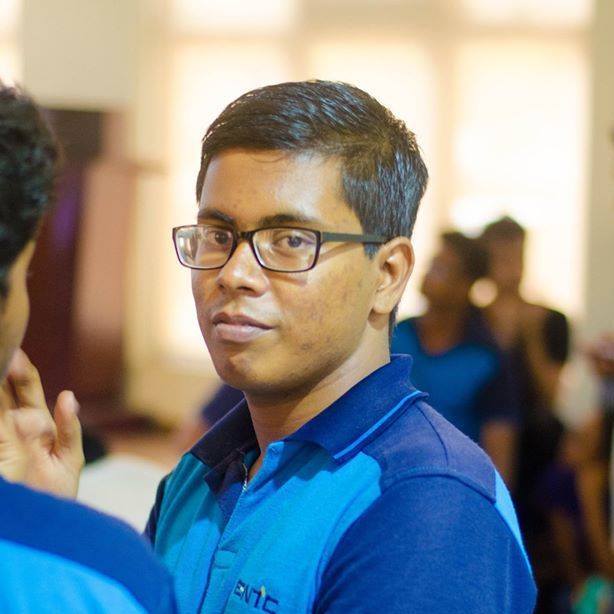}}]{Jathushan Rajasegran}
	received his Bachelors degree in Electronic and Telecommunication Engineering from University of Moratuwa, Sri Lanka in 2019. He is currently working as a Research Engineer at Inception Institute of Artificial Intelligence. He worked as a Visiting Research Intern at Data61-CSIRO, Sydney in 2017. His research interests include computer vision, pattern recognition and machine learning.
\end{IEEEbiography}

\vskip -3\baselineskip plus -1fil

\begin{IEEEbiography}[{\includegraphics[width=1in,height=1.25in,clip,keepaspectratio]{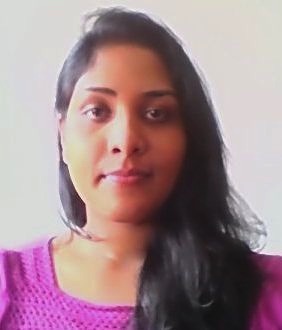}}]{Ashanie Gunathillake}
	received her PhD in Systems and Control from the University of New South Wales, Sydney, Australia, in 2018 and her B.Sc. degree in Electrical and Information Engineering from the University of Ruhuna, Sri Lanka, in 2011. She was a Research Assistant in University of Sydney in 2018. Her current research interests include wireless communication, navigation, machine learning and computer vision.
\end{IEEEbiography}

\vskip -3\baselineskip plus -1fil

\begin{IEEEbiography}[{\includegraphics[width=1in,height=1.25in,clip,keepaspectratio]{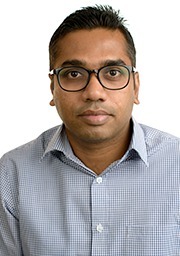}}]{Suranga Seneviratne}
is a Lecturer in Security at the School of Computer Science, The University of Sydney. He received his PhD from University of New South Wales, Australia in 2015. His current research interests include privacy and security in mobile systems, AI applications in security, and behaviour biometrics. Before moving into research, he worked nearly six years in the telecommunications industry in core network planning and operations. He received his bachelor degree from University of Moratuwa, Sri Lanka in 2005.
\end{IEEEbiography}
\vskip -2\baselineskip plus -1fil

\begin{IEEEbiography}[{\includegraphics[width=1in,height=1.25in,clip,keepaspectratio]{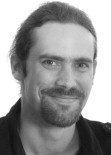}}]{Guillaume Jourjon}
	is senior researcher at Data61-CSIRO. He received his PhD from the University of New South Wales and the Toulouse University of Science in 2008. Prior to his PhD, he received a Engineer Degree from the ENSICA. He also received a DEUG in Physics and Chemistry (Major) and Mathematic (Minor) from the University of Toulouse III. His research areas of interest are related to Distributed Computing, Software Defined Network, in-Network Computing, and Security and Privacy of Networked Systems.
\end{IEEEbiography}
\newpage
\appendix
\setcounter{table}{0}
\renewcommand{\thetable}{A.\arabic{table}}

In this appendix, we describe additional experiments we conducted to evaluate the effect of using different pre-trained models, convolutional layers, and kernel functions to transform the feature space. In summary, our results are consistent and show the benefit of using style embeddings. Different layers and architectures show minor variations among themselves, however, none of them stands out as significantly better compared to others.  

\subsection{Appendix I}
\label{appendix:I}

\textcolor{black}{We used three different CNN architectures to compare the performance of neural embeddings; VGG19, VGG16 and ResNet50. We used the last fully connected layers (${fc\_7}$) $C \in \RR^{4096}$ of VGG19, VGG16 and the final average pooling layer (${avg\_pool}$) $C \in \RR^{2048}$ of ResNet50 as the content representations. To represent style, we tested and compared ten different convolution layer outputs of the three architectures. We considered the fifth convolution stage of each architecture to extract the Gram matrices. Extracted Gram matrices were reduced to $S \in \RR^{4096}$ using very sparse random projection. Details such as the kernel size, number of channels, and stride of the convolution layers used to extract Gram matrices for each architecture are shown in Table~\ref{Tab:ArchitectureDetails}.}

\textcolor{black}{In Table~\ref{Tab:PrecisionArchitectures} and Table~\ref{Tab:RecallArchitectures}, we present the results. Overall, performances of all architectures and the convolutional layers are approximately similar. Following high level observations can be made based on our results.} 

\begin{itemize}

    \item \textcolor{black}{In all four datasets, ResNet50 outperforms other two architectures in representing the content of app icons. For example, for all four k-NN scenarios, the content embeddings have approximately 1\%--3\% and 3\%--6\% higher performance in precision@k and recall@k in all apps dataset compared to other content representations.} \\ \vspace{-2mm}
    
    \item \textcolor{black}{Best performances for style representation vary among different layers of the three architectures for different datasets. For example, ResNet50 $conv5\_1$ shows the best performance for labelled set while VGG16 $conv5\_2$ shows the best performance for UKBench dataset for all k-NN scenarios outperforming other architectures by small margins.} \\ \vspace{-2mm}
    
    \item \textcolor{black}{For all apps dataset, VGG19 $conv5\_1$ outperforms the rest for 10-NN and 15-NN scenarios while VGG16 $conv5\_1$ shows the best performance for 5-NN and 20-NN (approximately 0.01\%--1\% higher performance in precision@k and recall@k than other architectures).} 
    
\end{itemize}

\begin{table}[h] 
\centering
\scriptsize
	\begin{tabular}{c|p{0.6cm}|p{1cm}|p{0.6cm}|p{0.75cm}|p{0.9cm}}
		\hline
		{\bf Layer} & {\bf Kernel Size} & {\bf No. of Channels} & {\bf Stride} & {\bf Padding} & {\bf Activation} \\ \hline
		VGG19 $conv5\_1$ & 3x3 & 512 & (1,1) & Same & ReLU\\ \hline
		VGG19 $conv5\_2$ & 3x3 & 512 & (1,1) & Same & ReLU\\ \hline
		VGG19 $conv5\_3$ & 3x3 & 512 & (1,1) & Same & ReLU\\ \hline
		VGG19 $conv5\_4$ & 3x3 & 512 & (1,1) & Same & ReLU\\ \hline
		VGG16 $conv5\_1$ & 3x3 & 512 & (1,1) & Same & ReLU\\ \hline
		VGG16 $conv5\_2$ & 3x3 & 512 & (1,1) & Same & ReLU\\ \hline
		VGG16 $conv5\_3$ & 3x3 & 512 & (1,1) & Same & ReLU\\ \hline
		ResNet50 $conv5\_1$ & 1x1 & 512 & (2,2)& Valid & ReLU\\ \hline
		ResNet50 $conv5\_2$ & 3x3 & 512 & (1,1)& Same & ReLU\\ \hline
		ResNet50 $conv5\_3$ & 1x1 & 2,048 & (2,2) & Valid & - \\
		\hline
	\end{tabular} \vspace{2mm}
	\caption{Convolution layer configurations}
	\label{Tab:ArchitectureDetails}
	\caption*{{\footnotesize{{\bf Note:} ResNet50 $conv5\_3$ does not have its own activation function as the output of $conv5\_3$ is added with a modified input tensor of the fifth convolution block before passing through a ReLU activation.}}}
	\vspace{-5mm}
\end{table}

\subsection{Appendix II}
\label{appendix:II}

\textcolor{black}{We next analyse the effect of transforming the Gram matrix into a different space. Earlier we used the style embeddings of the final convolution layer outputs of the VGG19 architecture and use the cosine distance to measure the similarity between the icons in the feature space. This approach might have a drawback, that there may be no degree of freedom available to tune the feature space specifically for the application, in this case the similarity of the icons. Hence, we transform the feature space into different kernel spaces, using parameterised kernel functions. This allows us to tune the hyper-parameters of the space to uniquely design a metric for our application. We experiment with polynomial kernel, squared exponential kernel~\cite{li2017demystifying}, and shifting the feature values to introduce a bias~\cite{novak2016improving}. The kernel transformations are given bellow.}

\begin{align}
    \textcolor{black}{POLY(a,b)(G) = (GG^T + a)^b}
\end{align}
\begin{align}
    \textcolor{black}{SHIFT(c)(F) = (F+c)(F+c)^T}
\end{align}
\begin{align}
    \textcolor{black}{SquaredExp(d)(G) = \exp^{(\frac{G}{d})}}
\end{align}

\textcolor{black}{Here, $G$ is the gram matrix and $F$ is the corresponding feature set. $a,b$ are the bias and polynomial coefficient of the polynomial kernel respectively. Parameter $c$ is the shifting bias of the $SHIFT(c)$ function, and $d$ is the variance of the Squared Exponential kernel.We show our results for the UKBench and Labelled Dataset in Table~\ref{Tab:Kernel:UKBench} and Table~\ref{Tab:Kernel:Labelled}, respectively. We make the following high level observations from our results}


\begin{itemize}
    \item \textcolor{black}{All three kernels performance is almost similar to the perceptual loss on feature space.}  \\ \vspace{-3mm}
    \item \textcolor{black}{Shifting the features by 1, gives a small gain in retrieval rates. However, smaller or larger values tends to decrease the performance. For example SHIFT(1) gives 81.25\% recall@5 value, while SHIFT(0.1) gives  81.27\% and SHIFT(10) gives 81.02\% on UKBench dataset. Therefore, it is necessary to find optimal values for the kernel hyper-parameters.} \\ \vspace{-3mm}
    \item \textcolor{black}{Polynomial and Squared Exponential kernels are more sensitive to the parameter selection. For Polynomial kernel, the power parameter is highly significant, as we can see from  Table~\ref{Tab:Kernel:UKBench} and Table~\ref{Tab:Kernel:Labelled}. For instance, POLY(x,2) has very low recall and precision compared to POLY(x,1). Further, the Exponential kernel has similar trends as SHIFT kernel, with respect to standard deviation, i.e. performance drops at lower and larger values of standard deviation. For example SquaredExp(1000) gives 82.24\% recall@5 value, while SquaredExp(100) and SquaredExp(10000) gives 35.96\% and 73.32\% respectively.} 
\end{itemize}

\begin{table*}  \centering 
	\scriptsize{
		\begin{tabular}{p{0.1cm}p{0.9cm}p{0.8cm}p{0.8cm}p{0.8cm}|p{0.8cm}p{0.8cm}p{0.8cm}p{0.8cm}|p{0.8cm}p{0.8cm}p{0.8cm}|p{0.8cm}p{0.8cm}p{0.8cm}}
		& & \multicolumn{3}{c}{\textbf{Content}}& \multicolumn{10}{c}{\textbf{Style}} \\
		& & VGG19 & VGG16 & ResNet50 & \multicolumn{4}{c}{VGG19} 
	& \multicolumn{3}{c}{VGG16} 
	& \multicolumn{3}{c}{ResNet50}  \\
	& & & & & $conv5\_1$ & $conv5\_2$ & $conv5\_3$ & $conv5\_4$ 
	& $conv5\_1$ & $conv5\_2$ & $conv5\_3$ 
	& $conv5\_1$ & $conv5\_2$ & $conv5\_3$  \\			
			\cmidrule[1pt]{2-15}
			& \bf{5-NN}             & 46.36 & 46.96 & 48.60 & 46.72  &  46.92 & 47.12 & 44.28 & 47.16& 46.88  & 45.36 & 47.92 & 46.68 & 48.00 \\
			& \bf{10-NN}               & 25.28 & 25.68 & 26.60 &25.24 &  25.16 & 25.34 & 24.24 & 25.46 & 25.24  & 24.70 & 25.84 & 25.50 & 25.86 \\
			
			& \bf{15-NN}             & 17.47 & 17.67 & 18.15 & 17.25 &  17.24 & 17.28 & 16.71 & 17.49 & 17.39 & 17.09 & 17.67 & 17.43 & 17.64 \\
			\rott{\rlap{\bf Holidays}}
			& \bf{20-NN} & 13.31 & 13.43 & 13.77 & 13.13 &  13.18 & 13.17 & 12.83 & 13.33 & 13.28 & 13.02 & 13.47 & 13.28 & 13.51  \\
			
			\cmidrule{2-15}
			
			& \bf{5-NN}             & 70.22 & 71.00 & 75.02 & 65.02 &  69.68 & 71.40 & 68.58 & 69.70 & 71.90  & 69.51 & 70.92 & 68.29 & 70.25 \\
			& \bf{10-NN}               & 36.90 & 37.02 & 38.67 & 33.86 &  36.12 & 36.91 & 36.13 & 36.20 & 37.26  & 36.37 & 36.84 & 35.69 & 36.56  \\
			
			& \bf{15-NN}              & 25.03 & 25.06 & 25.97 & 22.95  &  24.41 & 24.93 & 24.60 & 24.51 & 25.13  & 24.73 & 24.91 & 24.17 & 24.77 \\
			\rott{\rlap{\bf UKBench}}
			& \bf{20-NN} & 18.97 & 18.97 & 19.58 & 17.40 &  18.47 & 18.87 & 18.66 & 18.52 & 18.97  & 18.77 & 18.86 & 18.30 & 18.73  \\
			
			\cmidrule{2-15}
			
			& \bf{5-NN}           & 56.43 & 57.17 & 60.85 & 60.57  &  60.22 & 59.30 & 55.61 & 60.20 & 60.28  & 56.20 & 60.67 & 59.85 & 60.37  \\
			& \bf{10-NN}               & 33.69 & 33.78 & 35.87 & 35.84 &  35.45 & 34.81 & 32.56 & 35.48 & 35.16  & 32.76 & 35.89 & 35.09 & 35.45  \\
			
			& \bf{15-NN}              & 24.05 & 24.09 & 25.49 & 25.45  &  25.12 & 24.81 & 23.22 & 25.16 & 24.90  & 23.34 & 25.53 & 25.03 & 25.05  \\
			\rott{\rlap{\bf Labelled}}
			& \bf{20-NN} & 18.69 & 18.81 & 19.84 & 19.75  &  19.47 & 19.23 & 18.18 & 19.53 & 19.39  & 18.16 & 19.94 & 19.53 & 19.58  \\
			
			\cmidrule{2-15}
			
			& \bf{5-NN}            & 45.66 & 45.06 & 48.54 & 50.67  &  50.52 & 49.77 & 44.59 & 51.07 & 50.30  & 44.69 & 50.49 & 49.35 & 49.58  \\
			& \bf{10-NN}               & 26.08 & 25.88 & 28.29 & 29.65 &  29.29 & 28.83 & 25.38 & 29.58 & 29.22  & 25.66 & 29.45 & 28.33 & 28.60  \\
			\rott{\rlap{\bf All}}
			& \bf{15-NN}              & 18.35 & 18.26 & 20.12 & 21.00  &  20.79 & 20.40 & 17.94 & 20.91 & 20.58  & 18.03 & 20.86 & 20.12 & 20.48  \\
			
			& \bf{20-NN} & 14.18 & 14.16 & 15.62 & 16.15  & 16.05  & 15.76 & 13.83 & 16.18 & 15.87  & 13.89 & 16.12 & 15.68 & 15.88 \\
			
			\cmidrule[1pt]{2-15}

		\end{tabular}
		\caption{Precision@k for all test scenarios (NN* - Nearest Neighbours)} 
		\label{Tab:PrecisionArchitectures}}
\end{table*}

\begin{table*}  \centering 
	\scriptsize{
		\begin{tabular}{p{0.1cm}p{0.9cm}p{0.8cm}p{0.8cm}p{0.8cm}|p{0.8cm}p{0.8cm}p{0.8cm}p{0.8cm}|p{0.8cm}p{0.8cm}p{0.8cm}|p{0.8cm}p{0.8cm}p{0.8cm}}
		& & \multicolumn{3}{c}{\textbf{Content}}& \multicolumn{10}{c}{\textbf{Style}} \\
		& & VGG19 & VGG16 & ResNet50 & \multicolumn{4}{c}{VGG19} 
	& \multicolumn{3}{c}{VGG16} 
	& \multicolumn{3}{c}{ResNet50}  \\
	& & & & & $conv5\_1$ & $conv5\_2$ & $conv5\_3$ & $conv5\_4$ 
	& $conv5\_1$ & $conv5\_2$ & $conv5\_3$ 
	& $conv5\_1$ & $conv5\_2$ & $conv5\_3$  \\
			\cmidrule[1pt]{2-15}
			& \bf{5-NN}             & 77.73 & 78.74 & 81.49 & 78.34  &  78.67 & 79.01 & 74.25 & 79.07& 78.60  & 76.06 & 80.35 & 78.27 & 80.48 \\
			& \bf{10-NN}               & 84.78 & 86.12 & 89.20 &84.64 &  84.37 & 84.98 & 81.29 & 85.38 & 84.64  & 82.83 & 86.65 & 85.51 & 86.72 \\
			
			& \bf{15-NN}             & 87.86 & 88.87 & 91.28 & 86.79 &  86.72 & 86.92 & 84.04 & 87.99 & 87.46 & 85.98 & 88.87 & 87.66 & 88.73 \\
			\rott{\rlap{\bf Holidays}}
			& \bf{20-NN} & 89.27 & 90.07 & 92.35 & 88.06 &  88.40 & 88.33 & 86.05 & 89.40 & 89.07 & 87.32 & 90.34 & 89.07 & 90.61  \\
			
			\cmidrule{2-15}
			
			& \bf{5-NN}             & 87.78 & 88.75 & 93.78 & 81.27 &  87.10 & 89.25 & 85.73 & 87.13 & 89.87  & 86.89 & 88.65 & 85.36 & 87.81 \\
			& \bf{10-NN}               & 92.25 & 92.55 & 96.67 & 84.65 &  90.31 & 92.27 & 90.32 & 90.50 & 93.16  & 90.92 & 92.10 & 89.22 & 91.41  \\
			
			& \bf{15-NN}              & 93.85 & 93.99 & 97.38 & 86.08  &  91.52 & 93.47 & 92.24 & 91.91 & 94.22  & 92.72 & 93.41 & 90.65 & 92.87 \\
			\rott{\rlap{\bf UKBench}}
			& \bf{20-NN} & 94.83 & 94.86 & 97.91 & 86.98 &  92.34 & 94.35 & 93.30 & 92.59 & 94.86  & 93.83 & 94.29 & 91.49 & 93.65  \\
			
			\cmidrule{2-15}
			
			& \bf{5-NN}           & 64.26 & 65.10 & 69.29 & 68.97  &  68.58 & 67.53 & 63.32 & 68.55 & 68.64  & 64.00 & 69.09 & 68.15 & 68.75  \\
			& \bf{10-NN}               & 76.72 & 76.94 & 81.69 & 81.63 &  80.73 & 79.29 & 74.15 & 80.81 & 80.08  & 74.60 & 81.75 & 79.91 & 80.73  \\
			
			& \bf{15-NN}              & 82.17 & 82.31 & 87.09 & 86.95  &  85.82 & 84.74 & 79.32 & 85.96 & 85.08  & 79.74 & 87.20 & 85.50 & 85.59  \\
			\rott{\rlap{\bf Labelled}}
			& \bf{20-NN} & 85.14 & 85.67 & 90.39 & 89.94  &  88.70 & 87.57 & 82.82 & 88.98 & 88.30  & 82.71 & 90.82 & 88.98 & 89.18  \\
			
			\cmidrule{2-15}
			
			& \bf{5-NN}            & 51.99 & 51.31 & 55.27 & 57.70  &  57.53 & 56.68 & 50.78 & 58.15 & 57.28  & 50.89 & 57.50 & 56.20 & 56.46  \\
			& \bf{10-NN}               & 59.40 & 58.94 & 64.42 & 67.53 &  66.71 & 65.67 & 57.81 & 67.36 & 66.54  & 58.43 & 67.08 & 64.51 & 65.13  \\
			\rott{\rlap{\bf All}}
			& \bf{15-NN}              & 62.70 & 62.39 & 68.72 & 71.74  &  71.01 & 69.68 & 61.29 & 71.43 & 70.30  & 61.60 & 71.26 & 68.72 & 69.96  \\
			
			& \bf{20-NN} & 64.59 & 64.48 & 71.15 & 73.58  & 73.10 & 71.77 & 63.01 & 73.72 & 72.28  & 63.27 & 73.44 & 71.40 & 72.34 \\
			
			\cmidrule[1pt]{2-15}

		\end{tabular}
		\caption{Recall@k for all test scenarios (NN* - Nearest Neighbours)} 
		\label{Tab:RecallArchitectures}}
\end{table*}       

\setcounter{table}{0}
\renewcommand{\thetable}{B.\arabic{table}}

\setlength{\tabcolsep}{0.3cm}
\begin{table}
    \centering
        \begin{center}
        \tiny
        \begin{tabular}{p{1.2cm} | p{0.3cm} p{0.3cm} p{0.3cm} p{0.3cm} | p{0.3cm} p{0.3cm} p{0.3cm} p{0.3cm}}
        \toprule[0.4mm]
\multirow{2}{*}{Kernel} & \multicolumn{4}{c}{ Recall@k} & \multicolumn{4}{c}{ Precision@k} \\
& 5-NN & 10-NN & 15-NN & 20-NN & 5-NN & 10-NN & 15-NN & 20-NN  \\ \midrule
Perceptual loss 	 &      81.27	 &     84.65	 &     86.08	 &     86.98 	 &     65.02	 &     33.86	 &     22.95	 &     17.40 \\
POLY(0,1) 	 &      81.27	 &     84.65	 &     86.08	 &     86.98 	 &     65.02	 &     33.86	 &     22.95	 &     17.40 \\
POLY(0,2) 	 &       0.05	 &      0.10	 &      0.15	 &      0.20 	 &      0.04	 &      0.04	 &      0.04	 &      0.04 \\
POLY(0.1,1) 	 &      81.27	 &     84.65	 &     86.08	 &     86.98 	 &     65.02	 &     33.86	 &     22.95	 &     17.40 \\
POLY(0.5,1) 	 &      81.27	 &     84.65	 &     86.08	 &     86.98 	 &     65.02	 &     33.86	 &     22.95	 &     17.40 \\
POLY(1,1) 	 &      81.27	 &     84.65	 &     86.08	 &     86.98 	 &     65.02	 &     33.86	 &     22.95	 &     17.40 \\
POLY(-0.1,1) 	 &      81.27	 &     84.65	 &     86.08	 &     86.98 	 &     65.02	 &     33.86	 &     22.95	 &     17.40 \\
POLY(-0.5,1) 	 &      81.27	 &     84.65	 &     86.08	 &     86.98 	 &     65.02	 &     33.86	 &     22.95	 &     17.40 \\
POLY(-1,1) 	 &      81.27	 &     84.65	 &     86.08	 &     86.98 	 &     65.02	 &     33.86	 &     22.95	 &     17.40 \\
POLY(1,2) 	 &       0.05	 &      0.10	 &      0.15	 &      0.20 	 &      0.04	 &      0.04	 &      0.04	 &      0.04 \\
POLY(-1,2) 	 &       0.05	 &      0.10	 &      0.15	 &      0.20 	 &      0.04	 &      0.04	 &      0.04	 &      0.04 \\
SHIFT(0.1) 	 &      81.27	 &     84.63	 &     86.08	 &     86.98 	 &     65.02	 &     33.85	 &     22.95	 &     17.40 \\
SHIFT(0.2) 	 &      81.27	 &     84.62	 &     86.07	 &     86.97 	 &     65.02	 &     33.85	 &     22.95	 &     17.39 \\
SHIFT(0.5) 	 &      81.27	 &     84.64	 &     86.06	 &     86.96 	 &     65.02	 &     33.85	 &     22.95	 &     17.39 \\
SHIFT(1) 	 &      81.25	 &     84.59	 &     86.03	 &     86.90 	 &     65.00	 &     33.84	 &     22.94	 &     17.38 \\
SHIFT(2) 	 &      81.22	 &     84.58	 &     86.01	 &     86.88 	 &     64.97	 &     33.83	 &     22.94	 &     17.38 \\
SHIFT(5) 	 &      81.11	 &     84.60	 &     85.82	 &     86.74 	 &     64.89	 &     33.84	 &     22.89	 &     17.35 \\
SHIFT(10) 	 &      81.02	 &     84.44	 &     85.66	 &     86.41 	 &     64.82	 &     33.78	 &     22.84	 &     17.28 \\
SHIFT(15) 	 &      80.69	 &     84.23	 &     85.45	 &     86.21 	 &     64.55	 &     33.69	 &     22.79	 &     17.24 \\
SHIFT(20) 	 &      80.39	 &     83.83	 &     85.19	 &     86.07 	 &     64.31	 &     33.53	 &     22.72	 &     17.21 \\
SquaredExp(1) 	 &      25.89	 &     26.28	 &     26.59	 &     26.81 	 &     20.71	 &     10.51	 &      7.09	 &      5.36 \\
SquaredExp(10) 	 &      25.78	 &     26.86	 &     27.48	 &     27.77 	 &     20.63	 &     10.75	 &      7.33	 &      5.55 \\
SquaredExp(100) 	 &      35.96	 &     37.78	 &     38.57	 &     39.06 	 &     28.77	 &     15.11	 &     10.28	 &      7.81 \\
SquaredExp(500) 	 &      80.52	 &     84.36	 &     86.05	 &     87.20 	 &     64.42	 &     33.75	 &     22.95	 &     17.44 \\
SquaredExp(1000) 	 &      82.24	 &     85.80	 &     87.30	 &     88.28 	 &     65.79	 &     34.32	 &     23.28	 &     17.66 \\
SquaredExp(2000) 	 &      78.52	 &     81.91	 &     83.28	 &     84.30 	 &     62.82	 &     32.76	 &     22.21	 &     16.86 \\
SquaredExp(5000) 	 &      74.57	 &     77.57	 &     79.09	 &     80.02 	 &     59.65	 &     31.03	 &     21.09	 &     16.00 \\
SquaredExp(10000) 	 &      73.32	 &     76.35	 &     77.87	 &     78.78 	 &     58.66	 &     30.54	 &     20.77	 &     15.76 \\
SquaredExp(20000) 	 &      64.03	 &     68.48	 &     70.64	 &     72.12 	 &     51.22	 &     27.39	 &     18.84	 &     14.42 \\
\bottomrule[0.4mm]
\end{tabular}
\end{center}
\caption{Performance of kernel functions - UKBench}
\label{Tab:Kernel:UKBench}
\caption*{{\scriptsize{{\bf Note:} Results are based on the style embeddings from VGG19 $conv5\_1$. }}}
\end{table}

\setlength{\tabcolsep}{0.3cm}
\begin{table}
    \centering
        \begin{center}
        \tiny
        \begin{tabular}{p{1.2cm} | p{0.3cm} p{0.3cm} p{0.3cm} p{0.3cm} | p{0.3cm} p{0.3cm} p{0.3cm} p{0.3cm}}
        \toprule[0.4mm]
\multirow{2}{*}{Kernel} & \multicolumn{4}{c}{ Recall@k} & \multicolumn{4}{c}{ Precision@k} \\
& 5-NN & 10-NN & 15-NN & 20-NN & 5-NN & 10-NN & 15-NN & 20-NN  \\ \midrule
Perceptual loss 	 &      68.97	 &     81.63	 &     86.95	 &     89.94 	 &     60.57	 &     35.84	 &     25.45	 &     19.75 \\
POLY(0,1) 	 &      68.97	 &     81.63	 &     86.95	 &     89.94 	 &     60.57	 &     35.84	 &     25.45	 &     19.75 \\
POLY(0,2) 	 &       0.14	 &      0.28	 &      0.42	 &      0.57 	 &      0.12	 &      0.12	 &      0.12	 &      0.12 \\
POLY(0.1,1) 	 &      68.97	 &     81.63	 &     86.95	 &     89.94 	 &     60.57	 &     35.84	 &     25.45	 &     19.75 \\
POLY(0.5,1) 	 &      68.97	 &     81.63	 &     86.95	 &     89.94 	 &     60.57	 &     35.84	 &     25.45	 &     19.75 \\
POLY(1,1) 	 &      68.97	 &     81.63	 &     86.95	 &     89.94 	 &     60.57	 &     35.84	 &     25.45	 &     19.75 \\
POLY(-0.1,1) 	 &      68.97	 &     81.63	 &     86.95	 &     89.94 	 &     60.57	 &     35.84	 &     25.45	 &     19.75 \\
POLY(-0.5,1) 	 &      68.97	 &     81.63	 &     86.95	 &     89.94 	 &     60.57	 &     35.84	 &     25.45	 &     19.75 \\
POLY(-1,1) 	 &      68.97	 &     81.63	 &     86.95	 &     89.94 	 &     60.57	 &     35.84	 &     25.45	 &     19.75 \\
POLY(1,2) 	 &       0.14	 &      0.28	 &      0.42	 &      0.57 	 &      0.12	 &      0.12	 &      0.12	 &      0.12 \\
POLY(-1,2) 	 &       0.14	 &      0.28	 &      0.42	 &      0.57 	 &      0.12	 &      0.12	 &      0.12	 &      0.12 \\
SHIFT(0.1) 	 &      68.97	 &     81.63	 &     86.95	 &     89.94 	 &     60.57	 &     35.84	 &     25.45	 &     19.75 \\
SHIFT(0.2) 	 &      68.97	 &     81.63	 &     86.95	 &     89.94 	 &     60.57	 &     35.84	 &     25.45	 &     19.75 \\
SHIFT(0.5) 	 &      68.97	 &     81.58	 &     87.00	 &     89.94 	 &     60.57	 &     35.82	 &     25.47	 &     19.75 \\
SHIFT(1) 	 &      69.03	 &     81.61	 &     87.03	 &     89.94 	 &     60.62	 &     35.83	 &     25.48	 &     19.75 \\
SHIFT(2) 	 &      69.03	 &     81.58	 &     86.97	 &     89.94 	 &     60.62	 &     35.82	 &     25.46	 &     19.75 \\
SHIFT(5) 	 &      68.95	 &     81.52	 &     87.00	 &     89.94 	 &     60.55	 &     35.79	 &     25.47	 &     19.75 \\
SHIFT(10) 	 &      68.95	 &     81.63	 &     86.97	 &     90.19 	 &     60.55	 &     35.84	 &     25.46	 &     19.80 \\
SHIFT(15) 	 &      68.86	 &     81.63	 &     87.06	 &     90.19 	 &     60.47	 &     35.84	 &     25.48	 &     19.80 \\
SHIFT(20) 	 &      68.66	 &     81.66	 &     87.03	 &     90.08 	 &     60.30	 &     35.86	 &     25.48	 &     19.78 \\
SquaredExp(1) 	 &      27.15	 &     28.34	 &     29.19	 &     29.75 	 &     23.85	 &     12.44	 &      8.54	 &      6.53 \\
SquaredExp(10) 	 &      34.47	 &     38.12	 &     39.67	 &     40.58 	 &     30.27	 &     16.74	 &     11.61	 &      8.91 \\
SquaredExp(100) 	 &      43.23	 &     48.88	 &     51.31	 &     52.78 	 &     37.97	 &     21.46	 &     15.02	 &     11.59 \\
SquaredExp(500) 	 &      59.42	 &     69.51	 &     74.82	 &     78.16 	 &     52.18	 &     30.52	 &     21.90	 &     17.16 \\
SquaredExp(1000) 	 &      67.62	 &     80.19	 &     85.50	 &     89.09 	 &     59.38	 &     35.21	 &     25.03	 &     19.56 \\
SquaredExp(2000) 	 &      67.28	 &     79.71	 &     85.25	 &     88.39 	 &     59.08	 &     35.00	 &     24.95	 &     19.40 \\
SquaredExp(5000) 	 &      65.19	 &     76.72	 &     81.72	 &     84.49 	 &     57.25	 &     33.68	 &     23.92	 &     18.55 \\
SquaredExp(10000) 	 &      63.44	 &     74.57	 &     79.34	 &     81.92 	 &     55.71	 &     32.74	 &     23.23	 &     17.98 \\
SquaredExp(20000) 	 &      62.76	 &     73.27	 &     78.07	 &     80.79 	 &     55.11	 &     32.17	 &     22.85	 &     17.74 \\
\bottomrule[0.4mm]
\end{tabular}
\end{center}
\caption{Performance of kernel functions - Labelled}
\label{Tab:Kernel:Labelled}
\caption*{{\scriptsize{{\bf Note:} Results are based on the style embeddings from VGG19 $conv5\_1$. }}}
\end{table}

\end{document}